\documentclass[useAMS,usenatbib,usedcolumn]{mnras}

\pdfoutput=1
\usepackage{graphicx}
\usepackage{xcolor}
\usepackage{booktabs}
\usepackage{wrapfig}

\def\lesssim{\la}
\def\gtrsim{\ga}

\usepackage{flushend}

\usepackage{adjustbox}

\title[Velocity function of the local volume]{Constraining cosmology with the velocity function of low-mass galaxies}
\author[Schneider \& Trujillo-Gomez]{Aurel Schneider$^{1}$ and Sebastian Trujillo-Gomez$^{2}$
\\
{$^1$Institute for Astronomy, Department of Physics, ETH Zurich,
Wolfgang-Pauli-Strasse 27, 8093, Zurich, Switzerland}\\
{$^2$Institute for Computational Science, University of Zurich, Winterthurerstrasse 190, 8057 Zurich, Switzerland}\\
{Email: aurel.schneider@phys.ethz.ch}}

\begin{document}

\label{firstpage}
\maketitle

\begin{abstract}
The number density of field galaxies per rotation velocity, referred to as the velocity function, is an intriguing statistical measure probing the smallest scales of structure formation. In this paper we point out that the velocity function is sensitive to small shifts in key cosmological parameters such as the amplitude of primordial perturbations ($\sigma_8$) or the total matter density ($\Omega_{\rm m}$). Using current data and applying conservative assumptions about baryonic effects, we show that the observed velocity function of the Local Volume favours cosmologies in tension with the measurements from {\tt Planck} but in agreement with the latest findings from weak lensing surveys. While the current systematics regarding the relation between observed and true rotation velocities are potentially important, upcoming data from HI surveys as well as new insights from hydrodynamical simulations will dramatically improve the situation in the near future.  
\end{abstract}



\section{Introduction}
One of the main goals of cosmology is to determine the parameters of the standard model of cosmology, $\Lambda$-cold-dark-matter ($\Lambda$CDM), and to examine potential extensions of this minimal scenario. From an observational perspective, it is important to test $\Lambda$CDM using as many different probes as possible and covering a large variety of scales and redshifts.

The cosmic microwave background (CMB) has so far been the most important observable for cosmology, as it provides a clean and precise probe of the temperature fluctuation at large scales and very early times. CMB experiments such as {\tt COBE} \citep{Smoot:1992td}, {\tt WMAP} \citep{Spergel:2003cb}, and {\tt Planck} \citep{Ade:2013zuv} have transformed cosmology from a speculative into a precision science. However, the temperature fluctuations from the CMB are restricted to large cosmological scales and do not contain much information about the late-time epochs of the universe. It is therefore essential to consider other cosmological observables as well, which preferentially probe smaller scales and different cosmological epochs.

The most important current alternatives to the CMB are galaxy clustering and weak lensing measurements. Over the next years these probes are expected to open up a new era of cosmology. Several upcoming wide-field galaxy surveys such as {\tt DES} \citep{Abbott:2015swa}, {\tt Euclid} \citep{Laureijs:2011gra}, or {\tt LSST} \citep{Abell:2009aa} will provide unprecedented measurements of the lensing signal and the galaxy distribution.

Next to these prime cosmological probes, there are other ways to extract cosmological information out of the available astronomical data. Examples are supernova distance measurements \citep[e.g.][]{Riess:2006fw}, number counts of galaxy clusters \citep{Ade:2015fva}, or the Lyman-$\alpha$ forest \citep{McDonald:2004xn,Palanque-Delabrouille:2015pga}. The high complementarity of cosmological probes allows to test systematics and to break degeneracies between cosmological parameters.

The main goal of this paper is to show that the number of low-mass field galaxies as a function of their HI rotation velocity -- the galactic velocity function -- has the potential to become another promising observable for cosmology. The velocity function extends to the dwarf galaxy regime and therefore probes the smallest observable scales of structure formation well beyond the reach of the CMB, galaxy clustering, and weak lensing measurements. Upcoming HI surveys, such as the {\it Westerbork Northern Sky HI Survey} \citep[{\tt WNSHS};][]{Verheijen:2008np} and the {\it Widefield ASKAP L-band Legacy All-sky Blind surveY} \citep[{\tt WALLABY};][]{Duffy:2012gr} will provide a wealth of high-resolution kinematic data which will allow to significantly reduce the current observational uncertainties.

In the last decade, the first generation of large volume HI surveys such as {\tt HIPASS} \citep{Zwaan:2010aaa} and {\tt ALFALFA} \citep{Haynes:2011hi} have revealed a surprisingly shallow velocity function, apparently at odds with results from $N$-body simulations \citep{Zavala:2009ms,TrujilloGomez:2010yh,Papastergis:2011xe}. Motivated by these puzzling results, \citet[][henceforth K15]{Klypin:2014ira} performed a detailed analysis of the velocity function, studying the sample completeness of the galaxy catalogue by \citet{Karachentsev:2013ipr} and assuming a one-to-one relation between the observed HI profile half-width ($w_{50}$) and the maximum circular velocity ($v_{\rm max}$) of the galaxy's dark matter halo. They concluded that the discrepancy between theory and observations persists and cannot be attributed to the incompleteness of HI surveys.

Several papers have questioned the conclusions of K15, pointing out strong biases between $w_{50}$ and $v_{\rm max}$ observed in hydrodynamical simulations which could alleviate or completely solve the tension between theory and observations \citep{Brook:2015ofa,Maccio:2016egb,Brooks:2017rfe}. Recently, we performed a careful analysis of the velocity function including potential systematic effects from baryons \citep{Trujillo-Gomez:2016pix, Schneider:2016ayw}. We found that part but not all of the tension disappears when all uncertainties due baryonic processes are included and that the disagreement with results from hydrodynamical simulations could originate from subgrid implementations producing HI disks that are too small and turbulent.

However, all recent papers on the velocity function \citep[e.g.][]{Brook:2015ofa,Papastergis:2016aba,Maccio:2016egb,Trujillo-Gomez:2016pix,Schneider:2016ayw,Brooks:2017rfe} are based on the best-fitting cosmology of the {\tt Planck} CMB experiment. The {\tt Planck15} cosmology is characterised by both a large amplitude of perturbations ($\sigma_{8}$) and a large matter abundance ($\Omega_{\rm m}$), in tension with recent measurements based on weak lensing \citep{Hildebrandt:2016iqg} and cluster counts \citep[e.g.][]{Ade:2015fva}. In the present paper we investigate how the predicted galaxy velocity function depends on cosmological parameters. Surprisingly, we find that the tension between predicted and observed velocity function largely disappears for cosmologies closer to the best fitting values of weak lensing and cluster count experiments.

The paper is organised as follows: Sec.~\ref{VF} summarises our modelling approach; in Sec.~\ref{cosparams} we highlight the cosmology dependence of the velocity function, provide bounds on cosmological parameters ($\sigma_{8}$, $\Omega_{\rm m}$), and illustrate the effects of massive neutrinos and other hot dark matter components; Sec.~\ref{prospects} and \ref{conclusions} are dedicated to current limitations and future prospects of the velocity function as a cosmological probe. Finally, we provide additional information about different modelling steps in the Appendix.


\section{The velocity function of the Local Volume}\label{VF}
The number of galaxies per rotation velocity -- the velocity function -- is a powerful statistical probe, combining information about the abundance of galaxies with their density profiles. In this section we describe the observational data and discuss the analysis steps to determine the crucial relation between rotation velocity and maximum circular velocity. Furthermore, we summarise our method to obtain theoretical predictions including the treatment of non-gravitational effects from baryons. This section closely follows the method outlined in our previous work \citep{Trujillo-Gomez:2016pix,Schneider:2016ayw} to which we refer the reader for more detailed information. 

\subsection{Local Volume sample of galaxies}\label{LVsample}
The catalogue of \citet{Karachentsev:2013ipr} provides a complete galaxy sample of the Local Volume (i.e. within a distance of 10 Mpc from the Milky-Way) down to a luminosity of $M_B=-14$. Most of the galaxies from the Local Volume sample have unresolved HI 50 percent velocity profile line-widths ($w_{50}$). For the subset of late-type galaxies with no observable gas (less than ten percent of the full sample), the value of $w_{50}$ is estimated based on their K-band luminosities \citep[see][]{Trujillo-Gomez:2016pix}.

Based on HI line widths ($w_{50}$) and galaxy inclinations ($i$), K15 \citep{Klypin:2014ira} computed the Local Volume velocity function which they showed to be well fitted by the relation
\begin{equation}\label{obsvelfct}
\frac{dn}{d\ln v_{\rm rot}}=0.0782\left[\frac{100 \,\textrm{km/s}}{v_{\rm rot}}\right]\exp\left[-\left(\frac{v_{\rm rot}}{250 \,\textrm{km/s}}\right)^3\right]
\end{equation}
where the rotation velocity of the HI gas is defined as
\begin{equation}
v_{\rm rot}\equiv\frac{w_{50}}{2\sin(i)}.
\end{equation}

\begin{figure*}
\center{
\adjustbox{trim={0.02\width} {0.02\height} {0.02\width} {0.0\height},clip}{\includegraphics[width=.349\textwidth]{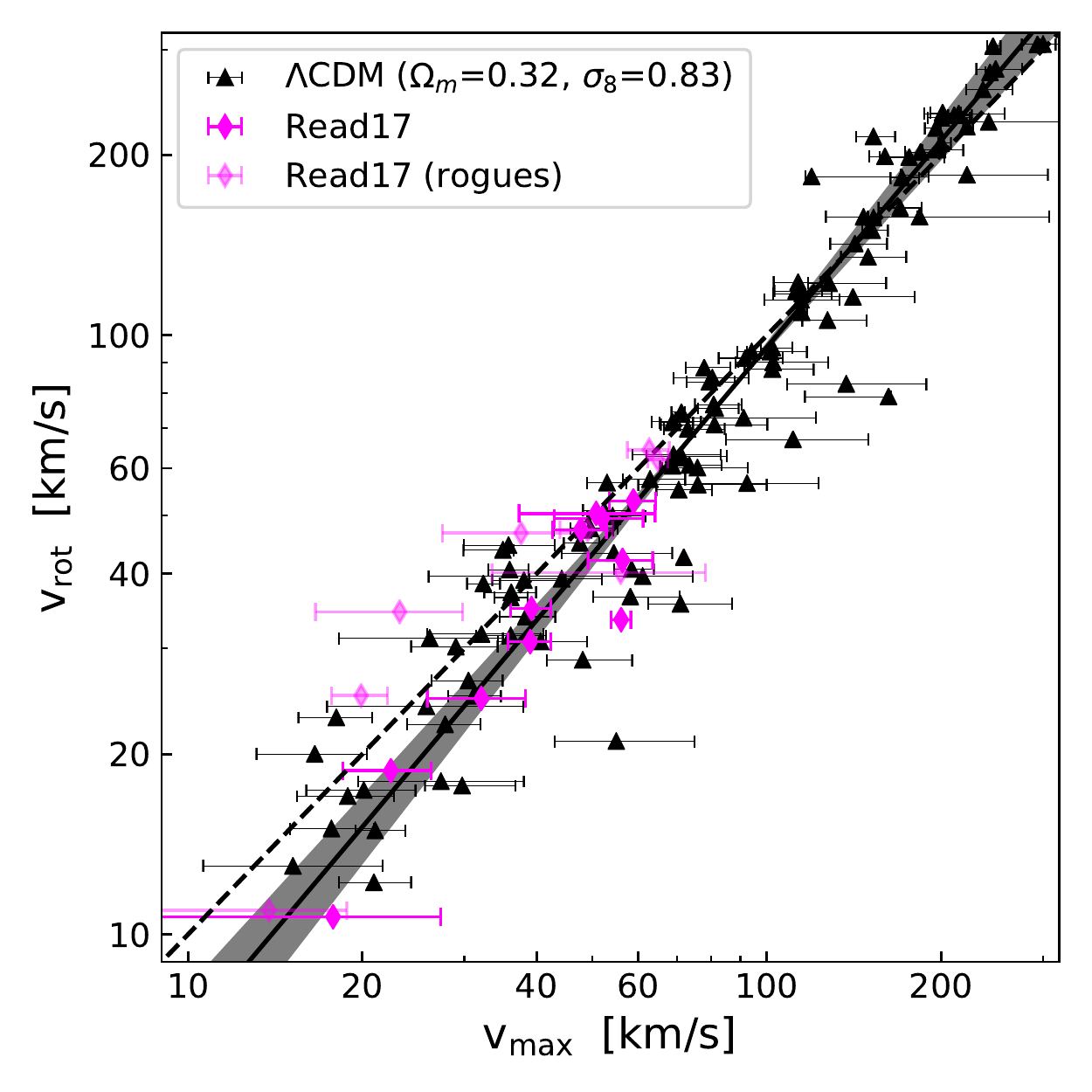}}
\adjustbox{trim={0.02\width} {0.02\height} {0.02\width} {0.0\height},clip}{\includegraphics[width=.349\textwidth]{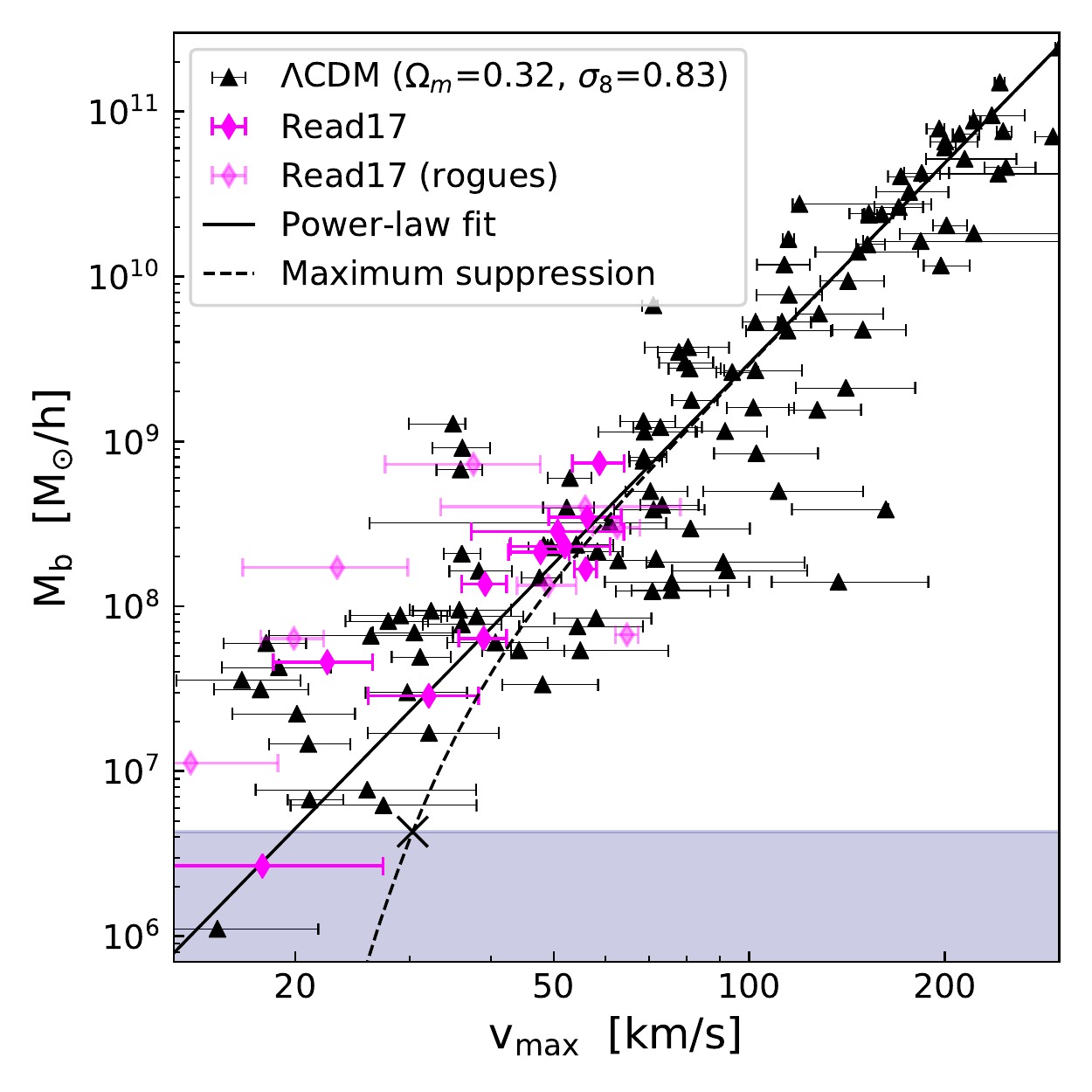}}
\adjustbox{trim={0.02\width} {0.02\height} {0.02\width} {0.0\height},clip}{\includegraphics[width=.314\textwidth]{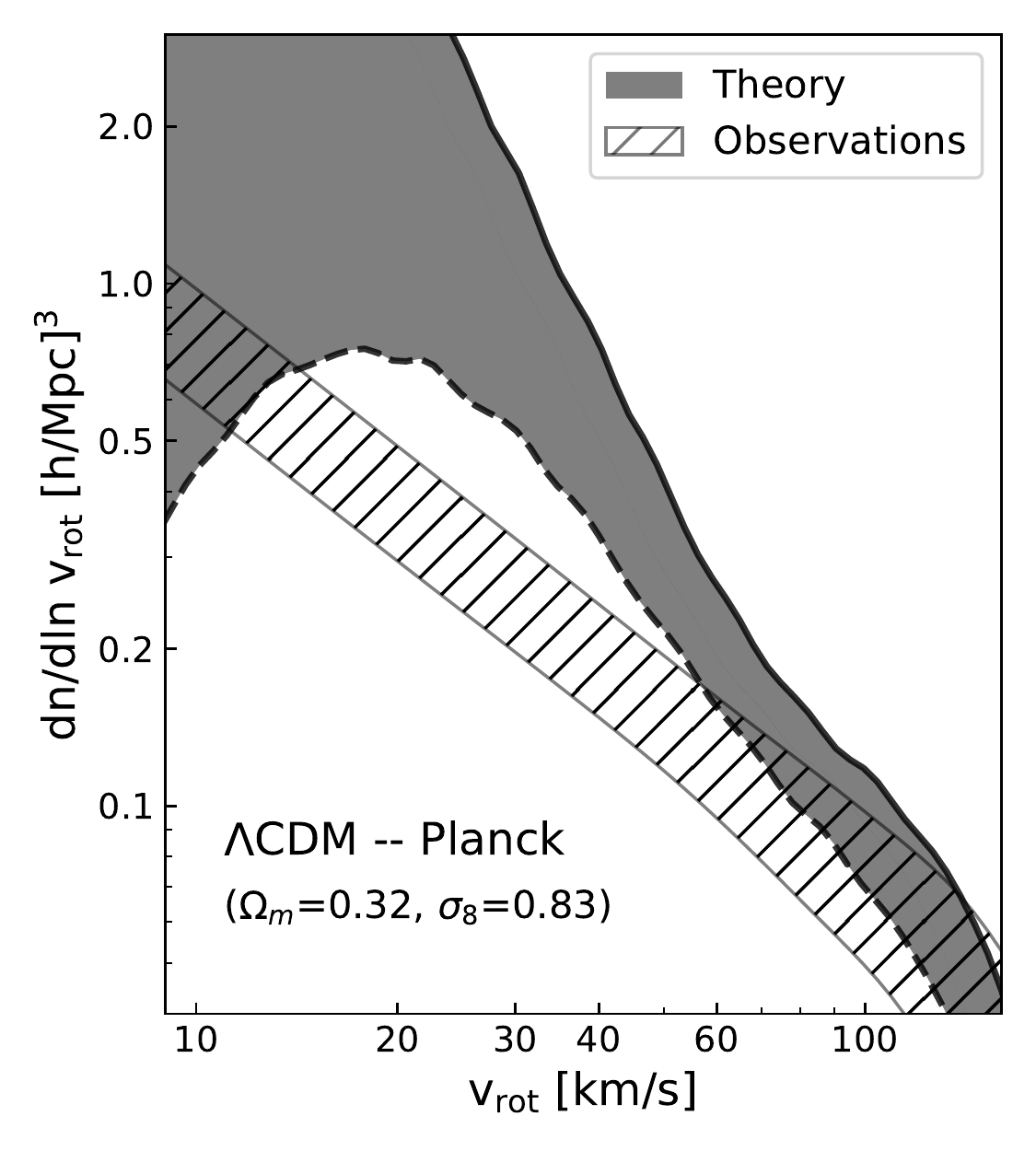}}
\caption{\label{fig:Planck}\emph{Left:} relation between rotation velocity ($v_{\rm rot}$) and maximum circular velocity ($v_{\rm max}$) of selected galaxies with spatially resolved kinematics (black data points). The best-fit power-law including the 3-$\sigma$ uncertainty of the fit is shown as solid line with grey-shaded band. For comparison, we add the results from Read17 \citep{Read:2016aaa} based on selected dwarf irregulars with high-quality rotation-curves (magenta symbols; the galaxies denoted \emph{rogues} potentially suffer from systematics, see Read17). \emph{Middle:} relation between baryonic mass ($M_{\rm b}$) and maximum circular velocity ($v_{\rm max})$ of the same galaxy sample. The solid line illustrates a simple power-law fit to the data, while the dashed line corresponds to a model featuring the maximum allowed suppression of baryonic mass due to photo-evaporation or supernova feedback (see text for more details). The data points from Read17 are again shown for comparison. \emph{Right:} predicted number of observable galaxies per rotation velocity including uncertainties due to baryonic effects (grey area) compared to the observed velocity function of the Local Volume \citep[hatched area,][]{Klypin:2014ira}. The rotation velocity ($v_{\rm rot}$) is defined via the HI line-width ($w_{50}$) and the galaxy inclination ($i$), i.e., $v_{\rm rot}\equiv w_{50}/(2\sin i)$. All three panels assume a $\Lambda$CDM universe with cosmological parameters from {\tt Planck}.}}
\end{figure*}

In order to use the Local Volume galaxy sample for cosmology, it is important to asses whether the Local Volume is a representative patch of the average universe. K15 performed a detailed analysis by searching for Local Volume analogues in the {\tt BolshoiP} simulations \citep{Klypin:2014kpa} and comparing their halo number densities to the total number densities of the {\tt BolshoiP} box\footnote{K15 defined Local Volume analogues to extend 10 Mpc around a central halo with $M=(1-2)\times 10^{12}$ M$_{\odot}$/h and to contain between 6 and 12 other haloes with $v_{\rm max}>170$ km/s. This definition is compatible with the observed properties of the Local Volume.}. On average, they found Local Volume analogues to have halo number densities very close to the universal one with a scatter between 25\% and 50\% slightly growing towards larger velocities. In the following, we use Eq.~(\ref{obsvelfct}) together with this scatter to reproduce the observed velocity function (see e.g. hatched bands in Figs. \ref{fig:Planck}, \ref{fig:VFcomp}, and \ref{fig:extensions}). In addition to the simulation study by K15, the velocity function has also been measured for larger volumes using blind HI surveys. The good agreement with results from {\tt HIPASS} and {\tt ALFALFA} confirms the representative nature of the Local Volume \citep[see e.g.][]{Klypin:2014ira,Papastergis:2014aba}.

\subsection{Selected galaxies and the connection between $v_{\rm rot}$ and $v_{\rm max}$}\label{vrotvmax}
The relation between the rotation velocity of gas ($v_{\rm rot}$) and the maximum circular velocity of the dark matter halo ($v_{\rm max}$) is of prime importance in order to compare observations with theory. It relates to the fundamental connection between observable galaxies and theoretically predicted properties of haloes. How strongly the values of $v_{\rm rot}$ deviate from $v_{\rm max}$ is currently debated especially at the scales of dwarf galaxies, with galaxy formation simulations often yielding contradictory results. Some recent results from hydrodynamical simulations suggest a strong bias that seems to originate from the highly turbulent and irregular HI distributions which arise due to strong stellar feedback \citep{Maccio:2016egb,Brooks:2017rfe}. Observationally based estimates from HI rotation curves, on the other hand, point towards a smaller difference between $v_{\rm rot}$ and $v_{\rm max}$ \citep{Read:2016aaa,Papastergis:2016aba,Trujillo-Gomez:2016pix}. The latter is further supported by ultra high-resolution simulations resolving the blast-waves of individual supernovae \citep{Read:2016bbb}.

In this paper, we build upon our past work \citep{Trujillo-Gomez:2016pix,Schneider:2016ayw} to estimate $v_{\rm max}$ using the largest available compilation of HI kinematic measurements of local field galaxies \citep[see][]{Papastergis:2016aba}\footnote{This compilation is based on the {\tt FIGGS} \citep{Begum:2008xc,Begum:2008gn}, {\tt THINGS} \citep{deBlok:2008wp,Oh:2010mc}, {\tt WHISP} \citep{Swaters:2009by,Swaters:2011yq}, {\tt LVHIS} \citep{Kirby:2012sg}, {\tt LITTLE THINGS} \citep{Hunter:2012un,Oh:2015xoa}, and {\tt SHIELD} \citep{Cannon:2011fe} galaxy samples. It furthermore contains galaxies from \citet{Sanders:1996ua}, \citet{Cote:2000aa}, \citep{Verheijen:2001ym}, and \citet{Trachternach:2009fb}. Finally it includes the recently discovered dwarf galaxy LeoP \citep{Giovanelli:2013maa,Bernstein-Cooper:2014tqa}.}. Each galaxy has at least one kinematic data point ($v_{\rm out}$) at the radius ($r_{\rm out}$), which is required to be more than three times further out than its half-light radius. This ensures the measurement to be well outside of the central region where feedback-induced cores may reduce the DM density compared to the pure NFW \citep{Navarro:1995iw} profile predicted by CDM  \citep{Read:2015sta}.

In order to estimate $v_{\rm max}$ for each galaxy, we fit an NFW profile through the data point ($r_{\rm out}$, $v_{\rm out}$) assuming an average halo concentration from a given concentration-mass relation. The resulting errors on $v_{\rm max}$ include the observational uncertainty in the measurement of $v_{\rm out}$ as well as the scatter of the concentration-mass relation for a given cosmology. 

Throughout this study, we only use one outermost data point ($r_{\rm out}$, $v_{\rm out}$) for the profile fitting, even for galaxies where more data is available. This allows us to develop a consistent pipeline for estimating $v_{\rm max}$ with galaxies from different surveys treated in exactly the same way. We want to point out that the majority of dwarf galaxies, which are the most crucial galaxies for this study, come from either {\tt FIGGS} or the {\tt SHIELD} sample with only one published kinematic data point \citep[see][for a more details]{Papastergis:2014aba}. In Appendix~\ref{app:read17} we further discuss the accuracy of the simplified fitting approach used in this paper.

The relation between $v_{\rm rot}$ and $v_{\rm max}$ from the selected galaxy sample can be approximated with a simple regression line of the form 
\begin{equation}\label{regressionline}
\ln v_{\rm max} = a\ln v_{\rm rot} + b
\end{equation}
where $a$ and $b$ are fitting parameters that depend on cosmology\footnote{We use simple least-square horizontal fits, as suggested by \citet{Trujillo-Gomez:2016pix}. We refer to this reference for a more detailed discussion regarding other techniques such as fits on previously binned data, weighted fits including error-bars, or vertical least-square fits.}. This simple approach provides a good description of the data points for all cosmologies considered in this paper.

In the left panel of Fig.~\ref{fig:Planck} we show the $v_{\rm rot}$-$v_{\rm max}$ relation for all galaxies of the selected sample (black symbols with error bars). The fit of Eq.~(\ref{regressionline}) is plotted as solid black line, where surrounding grey-shaded area indicates the 3-$\sigma$ uncertainties of the fitting parameters $a$ and $b$. We also show the one-to-one relation as a dashed line in order to illustrate the small deviation between $v_{\rm rot}$ and $v_{\rm max}$.

Next to our own results, we plot the $v_{\rm rot}$-$v_{\rm max}$ data points of a few nearby dwarf irregulars from \citet[][magenta symbols]{Read:2016aaa} which are based on high-quality rotation curves (instead of a single kinematic measurement), include the effects of baryon-induced cores, and use a more detailed correction for the pressure support. It is very encouraging that the results from \citet[][heceforth Read17]{Read:2016aaa} agree very well with our simplified analysis based on a larger sample of galaxies.

Six out of the nineteen data points from Read17 correspond to dwarfs that are also present in our selected galaxy sample. In Appendix~\ref{app:read17} we look at these galaxies individually and show that the $v_{\rm max}$-estimates from Read17 are well within the error bars obtained by our own analysis.

\subsection{Theoretical modelling of the velocity function}\label{model}
The velocity function is sensitive to both the abundance of haloes and the shape of the halo profile \citep{Zavala:2009ms,Schneider:2013wwa}. We will demonstrate in the following that it is the combination of these two dependencies that make the velocity function an excellent probe for cosmology. In this paper, we use a carefully calibrated theoretical model for the velocity function which was developed in \citet{Schneider:2016ayw}. The model is based on an extended Press-Schechter (EPS) approach \citep{Press:1973iz,Bond:1990iw} combined with estimates for the baryonic effects used in \citet{Trujillo-Gomez:2016pix} and \citet{Schneider:2016ayw}.

The halo number density as a function of mass is described by the equation
\begin{eqnarray}\label{massfct}
\frac{dn}{d\ln M}&=&\frac{1}{12\pi^2}\frac{\bar\rho}{M}\nu f(\nu)\frac{P_{\rm lin}(1/R)}{\delta_c^2R^3},\\
f(\nu)&=&A\sqrt{\frac{2\nu}{\pi}}(1+\nu^{-p}) {\rm e}^{-\nu/2},
\end{eqnarray}
where $P_{\rm lin}(k)$ is the linear power spectrum and where $\delta_c=1.686$, $A=0.322$, and $p=0.3$ \citep{Sheth:2001dp}. The variable $\nu= (\delta_c/\sigma)^2$ is obtained via the integral 
\begin{equation}
\sigma^2(R)=\int \frac{d\mathbf{k}^3}{(2\pi)^3} P_{\rm lin}(k)\Theta(1-kR),
\end{equation}
where halo mass ($M$) is connected to the filter scale ($R$) by the relation $M=4\pi\bar\rho(cR)^3/3$ with $c=2.5$ and where $\Theta$ is the Heaviside step function. Eq.~(\ref{massfct}) corresponds to the EPS halo mass function with sharp-$k$ filter that has been shown to give an accurate description for a large variety of cosmological scenarios \citep[][]{Schneider:2013ria,Schneider:2014rda,Benson:2012su}.

Obtaining the velocity function from the halo mass function requires information about the halo density profile. Here we assume an NFW profile which is a good approximation for the large majority of cosmological scenarios. This is at least true for the outer parts of the halo far away from potential DM halo cores that may result from astrophysical feedback mechanisms. Assuming an NFW profile results in the relation \citep[see e.g.][]{Sigad:2000cd} 
\begin{equation}\label{vmax}
v_{\rm max}=0.465\sqrt{\frac{GM}{r_{\rm vir}}}\left[c^{-1}\ln(1+c)-(1+c)^{-1}\right]^{-1/2}.
\end{equation}
Here, the concentration-mass relation $c\equiv c(M)$ is assumed to be a cosmology dependent stochastic function of halo mass. In general, halo concentrations have to be determined with the help of numerical simulations making the modelling of the cosmology dependence a difficult task.

In this paper, we use direct measurements of concentrations from $N$-body simulations whenever possible. If no simulations are available, we employ the empirical formula from \citet[henceforth D14]{Diemer:2014gba}, which is calibrated on a suite of $N$-body simulations with self-similar cosmologies and relates the value of the concentration to the effective slope of the power spectrum. The D14 model fully accounts for the cosmology dependence but it has the drawback of providing approximate results only. In order to obtain the most accurate results for a given cosmology $\chi$, we therefore apply the relation
\begin{equation}\label{concentration}
c(M|\chi) = \frac{c_{\rm D14}(M|\chi)}{c_{\rm D14}(M|\textrm{\tt Planck})} c(M|\textrm{\tt Planck})
\end{equation}
where $c_{\rm D14}$ is the concentration-mass relation from the D14 model, while
\begin{equation}
c(M|\textrm{\tt Planck})=10^{1.025}\left(\frac{10^{12}\, {\rm M_{\odot}/h}}{M}\right)^{0.097}
\end{equation}
is a direct fit to numerical simulations based on the {\tt Planck} cosmology \citep{Dutton:2014xda}. For the scatter of the concentrations at fixed halo mass we assume a cosmology independent value of 0.16 dex as suggested by D14. The approach of Eq.~(\ref{concentration}) guarantees very accurate concentrations for all cosmologies that are reasonably close to the parameters obtained by {\tt Planck} (i.e the region of parameter space we are most interested in). In Appendix~\ref{app:concentrations} we test this model of the concentration-mass relation against results from the literature assuming other cosmologies.

Based on a description for the mass function and the concentrations, it is now possible to calculate the $v_{\rm max}$ velocity function, i.e., the number density of haloes as a function of $v_{\rm max}$. Here we follow the method described in \citet{Schneider:2016ayw} which can be summarised as follows: (i) create a mock sample of haloes drawn from Eq.~(\ref{massfct}); (ii) assign random concentrations to each of the mock haloes assuming a Gaussian distribution with a mean given by  Eq.~(\ref{concentration}) and a standard deviation of 0.16 dex; (iii) calculate the corresponding values of $v_{\rm max}$ for each mock halo using Eq.~(\ref{vmax}).

For most of the cosmological scenarios, the scatter in the concentration-mass relation can safely be neglected. In this case the method simplifies considerably and the velocity function becomes
\begin{equation}
\frac{dn}{d\ln v_{\rm max}}=\frac{dn}{d\ln M}\frac{d\ln M}{d\ln v_{\rm max}},
\end{equation}
which means that it can be directly derived from the mass function using Eqs.~(\ref{vmax}) and (\ref{concentration}).

\subsection{Effects of baryonic physics on the velocity function}
The physical processes related to gas and star formation (including feedback) are expected to affect the evolution of DM haloes and the galaxies within them, modifying the predictions from gravity-only calculations. Here we follow the approach developed in \citet{Trujillo-Gomez:2016pix} and \citet{Schneider:2016ayw} which consists of constraining the maximum effect from baryons on the velocity function. We will now give a summary of all the baryonic effects and how they are modelled, referring the reader to the references above for more detailed information.

\subsubsection{Halo mass depletion from photo-evaporation and feedback}\label{depletion}
The observed baryon fraction of dwarf galaxies is much smaller than the average cosmic baryon fraction, $f_{\rm b}=\Omega_{\rm b}/\Omega_{\rm m}$, presumably due to a combination of stellar feedback gas ejection and UV photo-evaporation. The result of this gas ejection is a reduction of the halo mass (and consequently $v_{\rm max}$) compared to the value found in DM-only simulations.

The maximum baryon depletion can be estimated by assuming the entire baryon content to be evenly spread over the universe. In other words this means that the amplitude of perturbations ($\sigma_8$) is reduced to $\sigma_8\rightarrow (1-f_{\rm b})\sigma_8$. Assuming a hypothetical cosmology with non-clustering baryons leads to a shift of the maximum circular velocity of the order of
\begin{equation}\label{depl}
v_{\rm max}\rightarrow0.85\, v_{\rm max}
\end{equation}
in agreement with \citet{Schneider:2016ayw}\footnote{In \citet{Schneider:2016ayw} we assumed both a reduction of $\sigma_8$ and $\Omega_{\rm m}$ but we ignored how these changes affect the concentration-mass relation. The present estimate provides very similar results but is conceptually more consistent.}. A comparison with full hydrodynamical simulations \citep[e.g.][]{Sawala:2015cdf,Zhu:2015jwa} shows that Eq.~(\ref{depl}) is a good estimate for the maximum reduction of $v_{\rm max}$ due to baryonic physics.

\subsubsection{Suppression of observed galaxy abundance}\label{suppression}
The mechanisms ejecting baryons from galaxies do not only cause a reduction of $v_{\rm max}$, they also push some galaxies below the detection limit, making them unobservable for a given survey. This effect, which is mainly driven by photo-evaporation of cold gas and quenching of star formation at early times, leads to a scale dependent suppression in the observed abundance of low-$v_{\rm max}$ galaxies.

It is possible to estimate the maximum suppression effect by looking at the $M_{\rm b}$-$v_{\rm max}$ relation for dwarf galaxies. Any suppression of the baryonic content of galaxies below a certain mass would result in a downturn of the $M_{\rm b}$-$v_{\rm max}$ relation. Currently, such a downturn is not observed (see middle panel of Fig.~\ref{fig:Planck}), but we know from theory considerations that UV heating at the epoch of reionisation should induce a downturn at velocity scales below $v_{\rm max}\sim 20-30$ km/s \citep[e.g.][]{Gnedin:2000uj,Okamoto:2008sn}.

In order to find the maximum allowed downturn of the $M_{\rm b}$-$v_{\rm max}$ relation given the current uncertainties in the data, we assume a parametrisation of the form
\begin{equation}\label{Mbardownturn}
M_{\rm b, downturn}(v_{\rm max}|v_c) = \left[1+\left(\frac{v_c}{v_{\rm max}}\right)^4\right]^{-5}M_{\rm b}(v_{\rm max}),
\end{equation}
where $v_c$ is the characteristic scale of the downturn and $M_{\rm b}(v_{\rm max})$ is the linear fit to the data points\footnote{The relation of Eq.~(\ref{Mbardownturn}) corresponds to a subclass of the models studied in \citet{Trujillo-Gomez:2016pix}. It has a fixed shape for the suppression, which is motivated by recent results from hydrodynamical simulations \citep[e.g.][]{Sales:2016dmm}.}. With this parametrisation, it is now possible to define a maximum value of $v_c$ still allowed by the data using a \emph{maximum likelihood} analysis \citep[see][for more details]{Trujillo-Gomez:2016pix}.

As a next step we have to connect the $M_{\rm b}$-$v_{\rm max}$ relation to the velocity function. Since we know the detection limit of the full Local Volume galaxy sample to be $M_{\rm res}=4.3\times10^6$ M$_{\odot}$/h, the characteristic scale, where the velocity function is suppressed by a factor of two, can be estimated with Eq.~(\ref{Mbardownturn}) by solving
\begin{equation}\label{K13res}
M_{\rm b, downturn}(v_{s}|v_c)=M_{\rm res}
\end{equation}
for $v_{s}$. Assuming galaxies to be log-normally distributed around the $M_{\rm b}$-$v_{\rm max}$ relation (with a scatter $\sigma_s$), the suppression of the velocity function is then given by
\begin{equation}\label{suppr}
\frac{dn}{d\ln v_{\rm max}}\rightarrow \frac{1}{2}\left[{\rm erf}\left(\frac{\ln v_{\rm max} - \ln v_s}{ \sqrt{2}\ln\sigma_s}\right)+1\right]\frac{dn}{d\ln v_{\rm max}}.
\end{equation}
where $v_s$ defines characteristic suppression velocity. We refer the reader to \citet{Trujillo-Gomez:2016pix} for a more detailed discussion.

In the middle panel of Fig.~\ref{fig:Planck} we plot the $M_{\rm b}$-$v_{\rm max}$ data points of the selected galaxies together with the linear fit as solid black line. The model with maximum downturn allowed by the data (i.e. Eq.~\ref{Mbardownturn}) is shown as dashed line. The scale where it intersects the detection limit of the Local Volume sample (grey area) is highlighted by an $X$. This $X$ specifies the value of $v_s$, i.e. the characteristic scale of the downturn in the galactic velocity function.

\subsubsection{Dark matter cores from stellar feedback}
Another potential baryonic effect is the flattening of the inner density profiles of dwarf galaxies due to rapid fluctuations in the potential induced by efficient feedback gas blowouts \citep[e.g.][]{Mashchenko:2007jp,Governato:2012fa,DiCintio:2013qxa,Onorbe:2015ija,Chan:2015tna,Read:2015sta}. However, this effect is is already accounted for in our analysis (see Sec.~\ref{vrotvmax}). Due to the section criterion of our galaxy sample, all HI measurements lie well beyond the typical core radius \citep{Read:2015sta} in the regime where the density profile is well described by an NFW profile \citep[see][for a more detailed discussion]{Trujillo-Gomez:2016pix}.

\subsubsection{Stellar disk of massive galaxies}\label{velbias}
Finally, there is a potential baryonic effect that is completely negligible at dwarf galaxy scales but becomes relevant at larger velocities, where the fraction of baryons with respect to dark matter becomes more important. Beyond $v_{\rm max}\gtrsim100$ km/s, the baryonic mass of the galaxy dominates compared to the DM in the central region of the halo. This means that the effective $v_{\rm max}$ becomes larger than expected from gravity-only simulations \citep[see e.g.][]{Bekeraite:2016aaa}. Quantifying this effect is non-trivial as it depends on the details of how stars and gas are distributed within a galaxy as a function of its halo mass. 

In this paper we are predominantly interested in velocity range below $v_{\rm max}\sim100$ km/s where this effect is subdominant. We nevertheless perform a correction of the form
\begin{equation}\label{vmaxbias}
v_{\rm max} \rightarrow  \left[\frac{\alpha y^{6}}{1+y^{8}}+1\right]v_{\rm max},\hspace{0.5cm}y=\frac{v_{\rm max}}{v_{\alpha}}
\end{equation} 
where $v_{\alpha}=125$ km/s and $\alpha=0.2,0.4,0.5$ for the minimum, average, and maximum effect on $v_{\rm max}$. Eq.~(\ref{vmaxbias}) is based on the analysis described in Appendix \ref{app:bias}, where we quantify the effect of the observed stellar density profiles on the total circular velocity profile of the galaxy\footnote{We have checked that the effect from the gas component is negligible, see Appendix~\ref{app:bias}.}. The stellar density profiles are assumed to follow an exponential curve with structural parameters obtained by observations from SDSS \citep[see Appendix \ref{app:bias} and][]{Dutton:2010pz}. We furthermore correct for the adiabatic contraction of the inner DM distribution, an effect induced by the presence of a dominant central stellar component. Depending on the stellar density profiles and the model for the adiabatic contraction (AC), different results can be obtained. The minimum, average, and maximum effect refer to models with no AC, realistic AC (calibrated on $N$-body results), and strong AC (based on angular momentum conservation). It is expected that the real effect from adiabatic contraction has to lie between the minimum ($\alpha=0.2$) and maximum ($\alpha=0.5$) model. In principle, the effect on $v_{\rm max}$ is mildly cosmology dependent, a fact that we ignore for simplicity (see Appendix \ref{app:bias} for more details about the model).

\subsection{Putting everything together: the $v_{\rm rot}$ velocity function}\label{velfctvrot}
So far, we have described a theoretical model for the velocity function as a function of $v_{\rm max}$. In order to directly compare theory with observations, however, it is more useful to use the directly observable variable $v_{\rm rot}$, i.e. the deprojected line-of-sight profile width. This can be achieved with the simple transformation
\begin{equation}\label{VFvrot}
\frac{dn}{d\ln v_{\rm rot}}=\frac{dn}{d\ln v_{\rm max}}\frac{d\ln v_{\rm max}}{d\ln v_{\rm rot}}
\end{equation}
based on the relation of Eq.~(\ref{regressionline}). In principle, it would be possible to also include a scatter of the $v_{\rm max}$-$v_{\rm rot}$ relation. However, for the sample of selected galaxies with spatially resolved kinematics the observed scatter is similar to the size of the error-bars, which are dominated by the scatter in concentrations  that is already included in the analysis. We therefore follow the simple transformation of Eq.~(\ref{VFvrot}) without adding additional scatter.

So far, we have investigated the maximum and the minimum influence of baryons on the velocity function, discussing effects such as the depletion of $v_{\rm max}$, the suppression of observable galaxies, the potential core creation from violent feedback mechanisms, or the increase of $v_{\rm max}$ due to stars in high-mass galaxies. In the following, we add up all effects and present both an upper and a lower limit for the velocity function. This allows us to quantify our ignorance in terms of baryon physics and to provide a prediction including an uncertainty range. We now give a short summary of the main steps to obtain the velocity function for these two extreme cases.

The \emph{upper limit} of the velocity function is modelled ignoring both the depletion of $v_{\rm max}$ (Sec.~\ref{depletion}) and the suppression of the observable galaxy abundance (Sec.~\ref{suppression}) but including the increase of $v_{\rm max}$ of high-mass galaxies (i.e. Eq.~\ref{vmaxbias}, with $v_{\alpha}=125$ km/s and $\alpha=0.5$). The $v_{\rm rot}$-$v_{\rm max}$ relation is modelled with  Eq.~(\ref{regressionline}) using parameters $a$ and $b$ that are 3-$\sigma$ away from their best-fitting values in order to \emph{minimise} the difference between $v_{\rm rot}$ and  $v_{\rm max}$ at small velocities. Finally, the upper limit for the $v_{\rm rot}$ velocity function is obtained using Eq.~(\ref{VFvrot}).

The \emph{lower limit} of the velocity function is modelled including both the maximally allowed depletion and suppression effects but using the minimum model for the increase of $v_{\rm max}$ in high-mass galaxies. First of all, $v_{\rm max}$ is modified according to Eq.~(\ref{depl}) and Eq.~(\ref{vmaxbias}) with $v_{\alpha}=125$ km/s and $\alpha=0.2$. The maximum suppression of the observed galaxy abundance is then included using Eq.~(\ref{suppr}) with $v_c$ obtained via Eq.~(\ref{K13res}) and based on a \emph{maximum likelihood} analysis. We furthermore use Eq.~(\ref{regressionline}) to model the $v_{\rm rot}$-$v_{\rm max}$ relation, with $a$ and $b$ parameters that are 3-$\sigma$ away from their best-fitting values, \emph{maximising} the difference between $v_{\rm rot}$ and  $v_{\rm max}$ at small velocities. Finally, we again use Eq.~(\ref{VFvrot}) to obtain the lower limit of the $v_{\rm rot}$ velocity function.

In the right panel of Fig.~\ref{fig:Planck} we illustrate the lower and upper limits of the velocity function assuming the {\tt Planck} cosmology (dashed and solid black lines, respectively). The grey shaded area between these extreme cases quantifies the current uncertainty related to poorly understood baryonic effects. At scales above $v_{\rm rot}\sim50$ km/s, the uncertainties mostly come from the depletion effect of $v_{\rm max}$ and the potential inaccuracies from the $v_{\rm rot}$-$v_{\rm max}$ fitting relation. Below $v_{\rm rot}\sim50$ km/s, on the other hand, the uncertainties increase dramatically towards small velocities, which is mainly because of the lack of kinematic data from very faint galaxies that is needed to constrain the effects of high-redshift photo-evaporation and its coupling to feedback. 

The right panel of Fig.~\ref{fig:Planck} also shows the observed velocity function of the Local Volume as a hatched band, the width of the band indicating the uncertainty due to sampling variance of the Local Volume and statistical uncertainties (see discussion in Sec.~\ref{LVsample}). Despite maximising baryonic effects, a tension remains between theory and observation in agreement with our previous findings \citep{Trujillo-Gomez:2016pix,Schneider:2016ayw}.

\begin{figure*}
\center{
\adjustbox{trim={0.03\width} {0.02\height} {0.02\width} {0.0\height},clip}{\includegraphics[width=.255\textwidth]{Figs/VF_Planck.pdf}}
\adjustbox{trim={0.03\width} {0.02\height} {0.02\width} {0.0\height},clip}{\includegraphics[width=.255\textwidth]{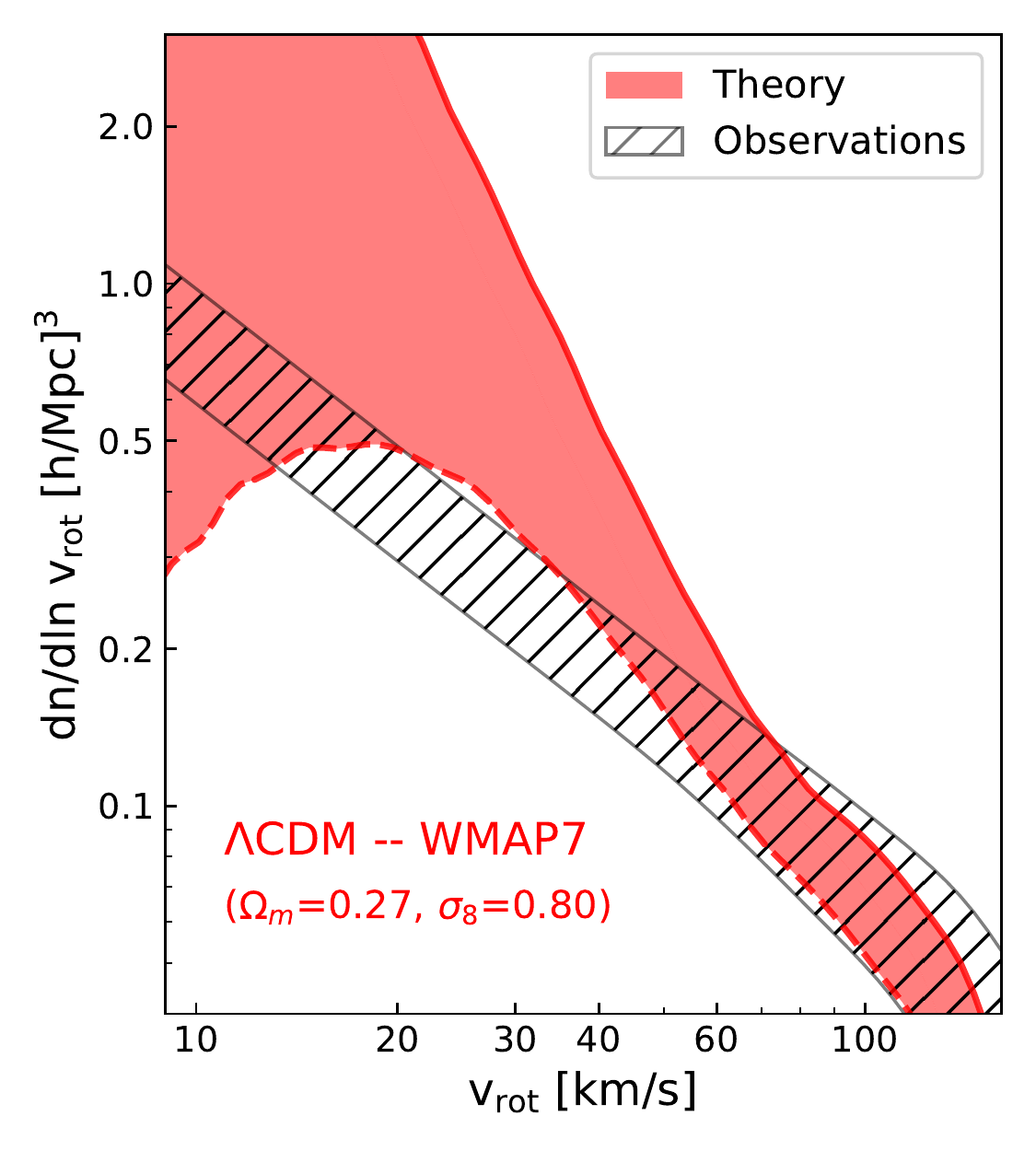}}
\adjustbox{trim={0.03\width} {0.02\height} {0.02\width} {0.0\height},clip}{\includegraphics[width=.255\textwidth]{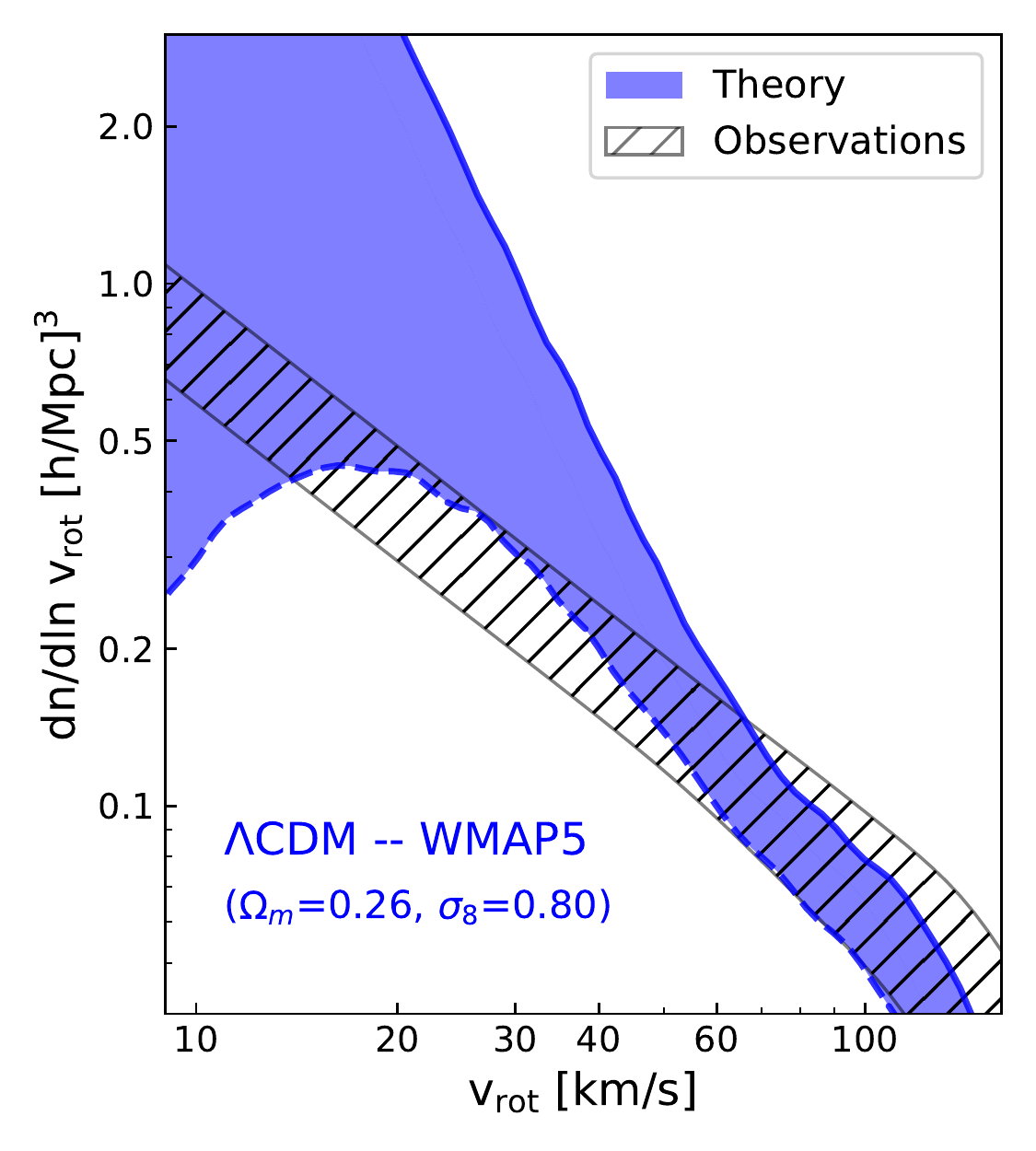}}
\adjustbox{trim={0.03\width} {0.02\height} {0.02\width} {0.0\height},clip}{\includegraphics[width=.255\textwidth]{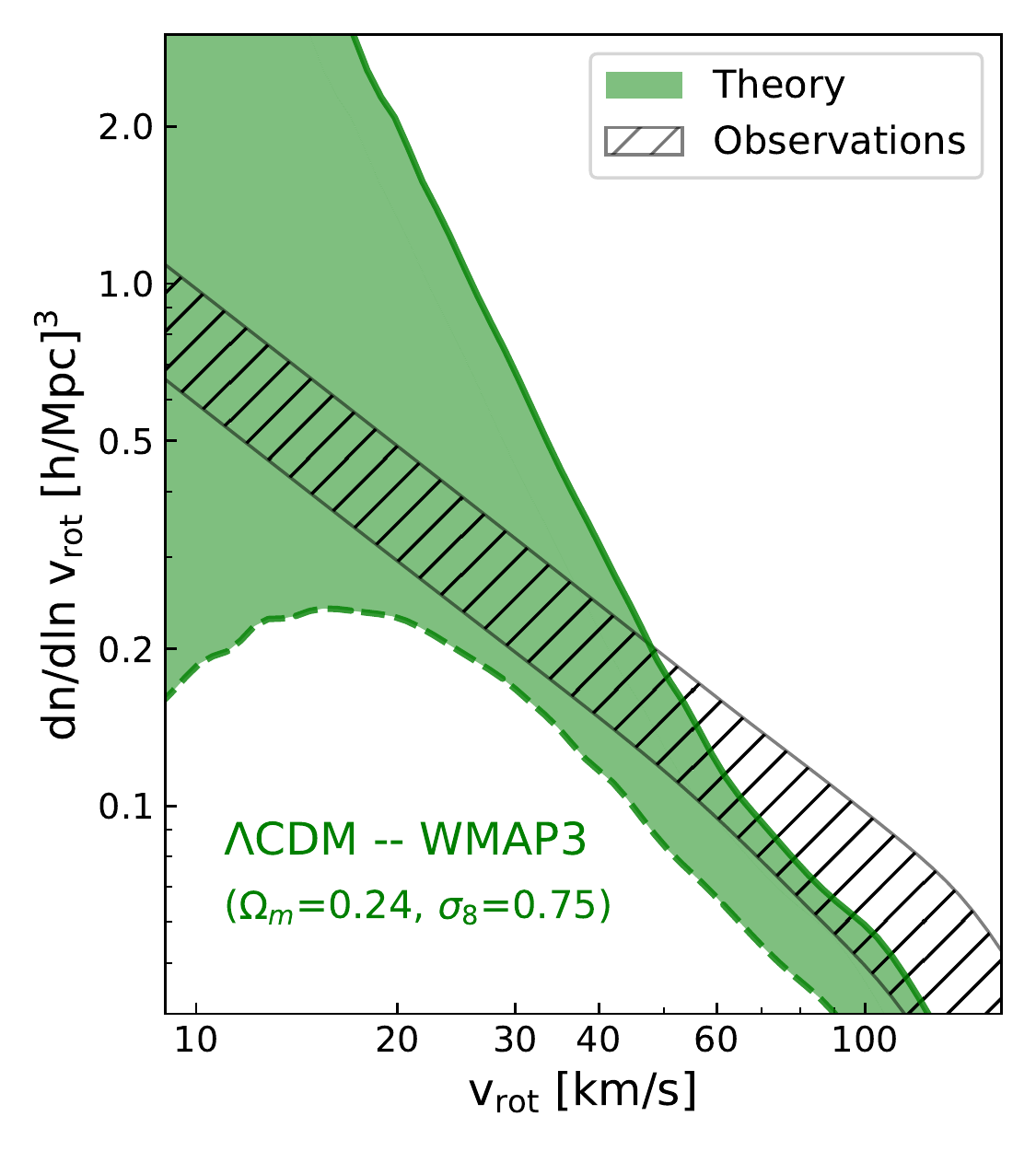}}
\caption{\label{fig:VFcomp}Velocity function for four different cosmologies based on {\tt Planck} (black), {\tt WMAP7} (red), {\tt WMAP5} (blue), and {\tt WMAP3} (green). The hatched areas correspond to the observed velocity function of the Local Volume including sample variance \citep{Klypin:2014ira}. The colour-shaded areas show the theoretical predictions using our model to account for the effects of baryons. The solid and dashed lines enclosing the coloured areas represent the minimum and maximum baryonic modifications allowed by current data (see Sec.~\ref{VF} for details). The rotation velocity ($v_{\rm rot}$) is defined via the HI line-width ($w_{50}$) and the galaxy inclination ($i$), i.e., $v_{\rm rot}\equiv w_{50}/(2\sin i)$.}}
\end{figure*}

\section{Cosmology dependence of the galaxy velocity function}\label{cosparams}
The standard model of cosmology, $\Lambda$CDM, is described by a number of parameters including the abundances of energy components ($\Omega_{\rm b}$, $\Omega_{\rm m}$, $\Omega_{\Lambda}$), the amplitude of fluctuations ($\sigma_8$), the spectral index ($n_s$), the Hubble constant ($H_0$), or the mass of the neutrino flavour states ($m_{\nu,i}$, $i=1,2,3$). In this section we study the sensitivity of the theoretically predicted velocity function of $\Lambda$CDM to some of the most important parameters (such as $\Omega_{\rm m}$, $\sigma_8$, $m_{\nu, i}$) and we identify parts of the cosmological parameter space where theory agrees with current observations.

\subsection{Velocity function for {\tt Planck} and {\tt WMAP} cosmologies}\label{VFcomp}
In order to highlight the sensitivity of the velocity function to small changes in cosmological parameters, we start with a comparison of the best-fitting cosmologies from {\tt Planck}, {\tt WMAP7}, {\tt WMAP5}, and {\tt WMAP3}. These cosmologies are well studied in the literature, allowing us to cross check some of our results (see also Appendix~\ref{app:EPStest} and \ref{app:concentrations}). Furthermore, the {\tt WMAP} parameters are in good agreement with the best fitting cosmologies form recent weak lensing shear \citep{Heymans:2013fya,Abbott:2015swa,Hildebrandt:2016iqg} and galaxy cluster probes \citep{Mantz:2014paa,Ade:2015fva}. This means that we can put the velocity function into context with the detected tension between the latest CMB and large-scale structure probes.

For the cosmologies studied in this section, we do not rely on the estimate for the concentrations based on Eq.~(\ref{concentration}) but instead use direct measurements from $N$-body simulations \citep{Maccio:2008pcd,Dutton:2014xda}. This allows us to obtain as accurate predictions as possible for the velocity function. Later on when studying general cosmologies (see Secs.~\ref{WL}-\ref{altDM}), we will be forced to use an approximation for the concentration-mass relation.

In Fig.~\ref{fig:VFcomp} we plot the observed velocity function of the Local Volume (hatched black band, K15) together with the theoretical predictions for the {\tt Planck}, {\tt WMAP7}, {\tt WMAP5}, and {\tt WMAP3} cosmologies (from left to right). As mentioned earlier, the velocity function based on {\tt Planck} is in disagreement with observations at small velocities below $v_{\rm rot}\sim 60$ km/s. For the {\tt WMAP} cosmologies, however, the disagreement is strongly reduced. This is most notable for the cases of {\tt WMAP5} and {\tt WMAP3}, where the error contours from theory and observations overlap at all scales. Intriguingly, the {\tt WMAP5} cosmology is in good agreement with the recent results from weak lensing shear {\tt CFHTlens} \citep{Heymans:2013fya} and {\tt KiDS} \citep{Hildebrandt:2016iqg}, as we will discuss in Sec.~\ref{WL}.

While the agreement between observations and theory is much better for {\tt WMAP} than for {\tt Planck}, the slope of the observed velocity function is still somewhat shallower than what the predictions suggest, at least for the {\tt WMAP7} and {\tt WMAP5} cosmologies. This could be either due to underestimated biases in the rotation-curve fitting \citep[see e.g.][]{Verbeke:2017rfd} or, more speculatively, due to new physics. One obvious example for the latter is a modified DM sector, as discussed in \citet[see also Sec.~\ref{altDM}]{Schneider:2016ayw}.

\begin{figure*}
\center{
\adjustbox{trim={.03\width} {0.02\height} {0.02\width} {0.0\height},clip}{\includegraphics[width=.255\textwidth]{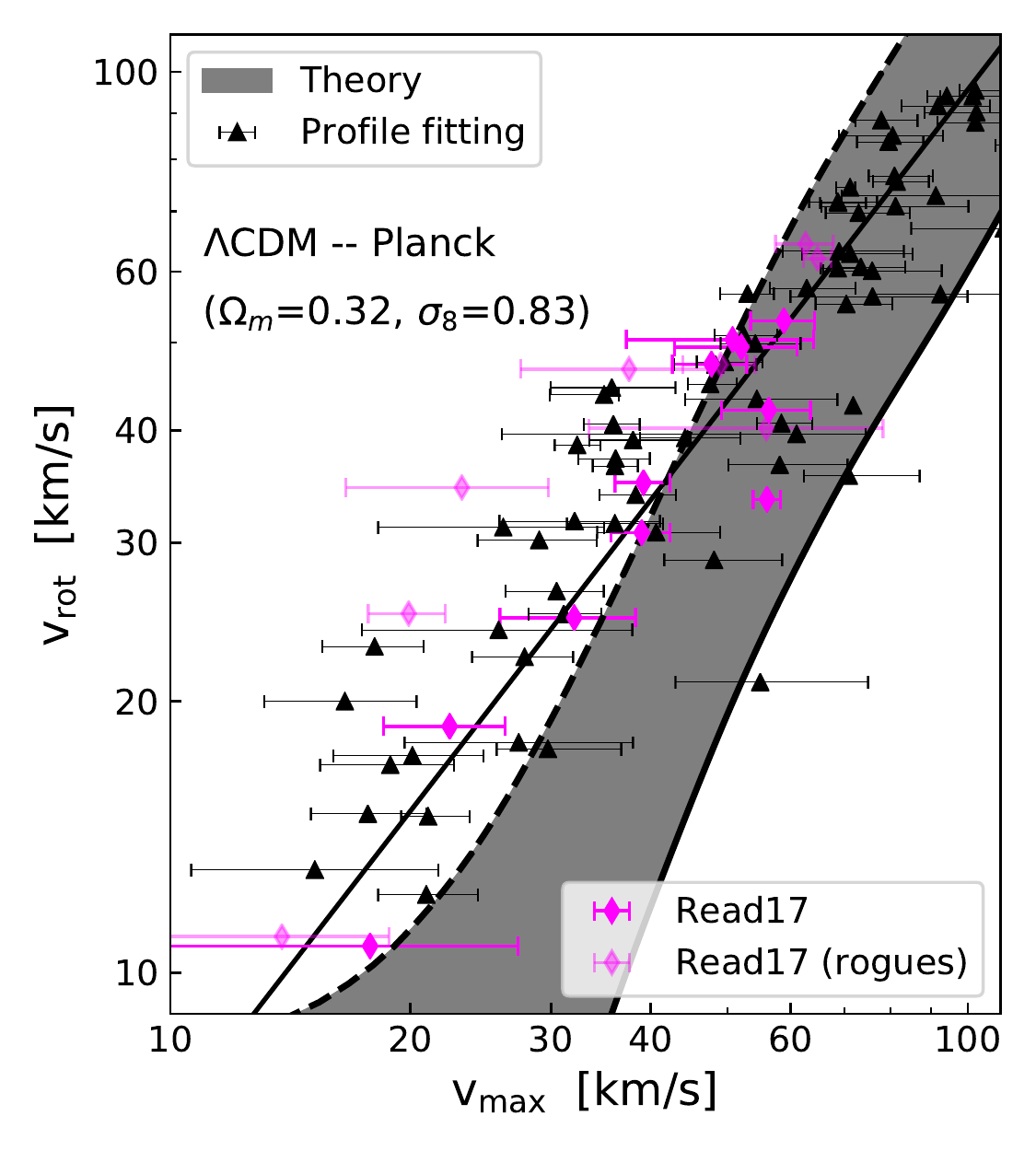}}
\adjustbox{trim={.03\width} {0.02\height} {0.02\width} {0.0\height},clip}{\includegraphics[width=.255\textwidth]{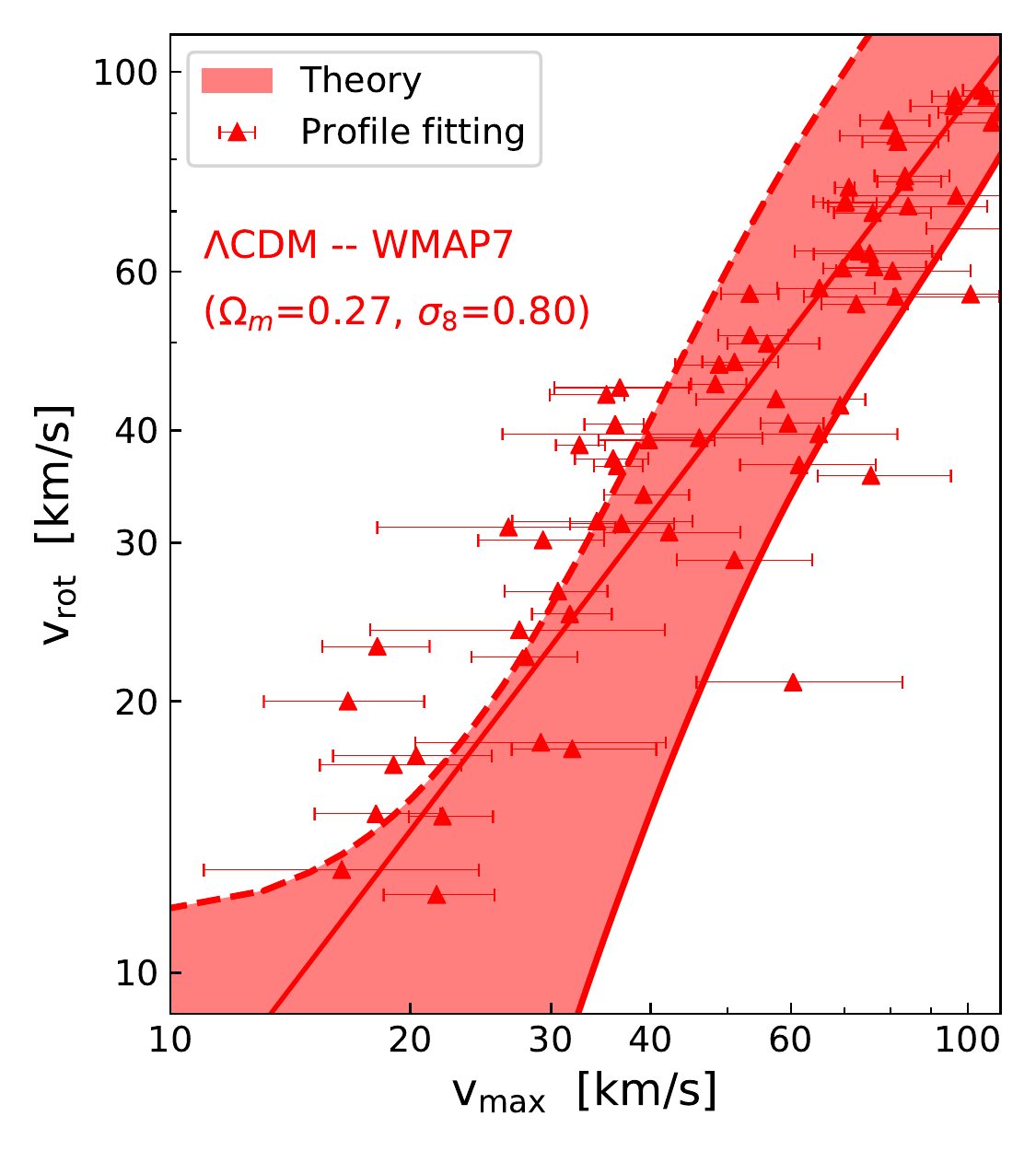}}
\adjustbox{trim={.03\width} {0.02\height} {0.02\width} {0.0\height},clip}{\includegraphics[width=.255\textwidth]{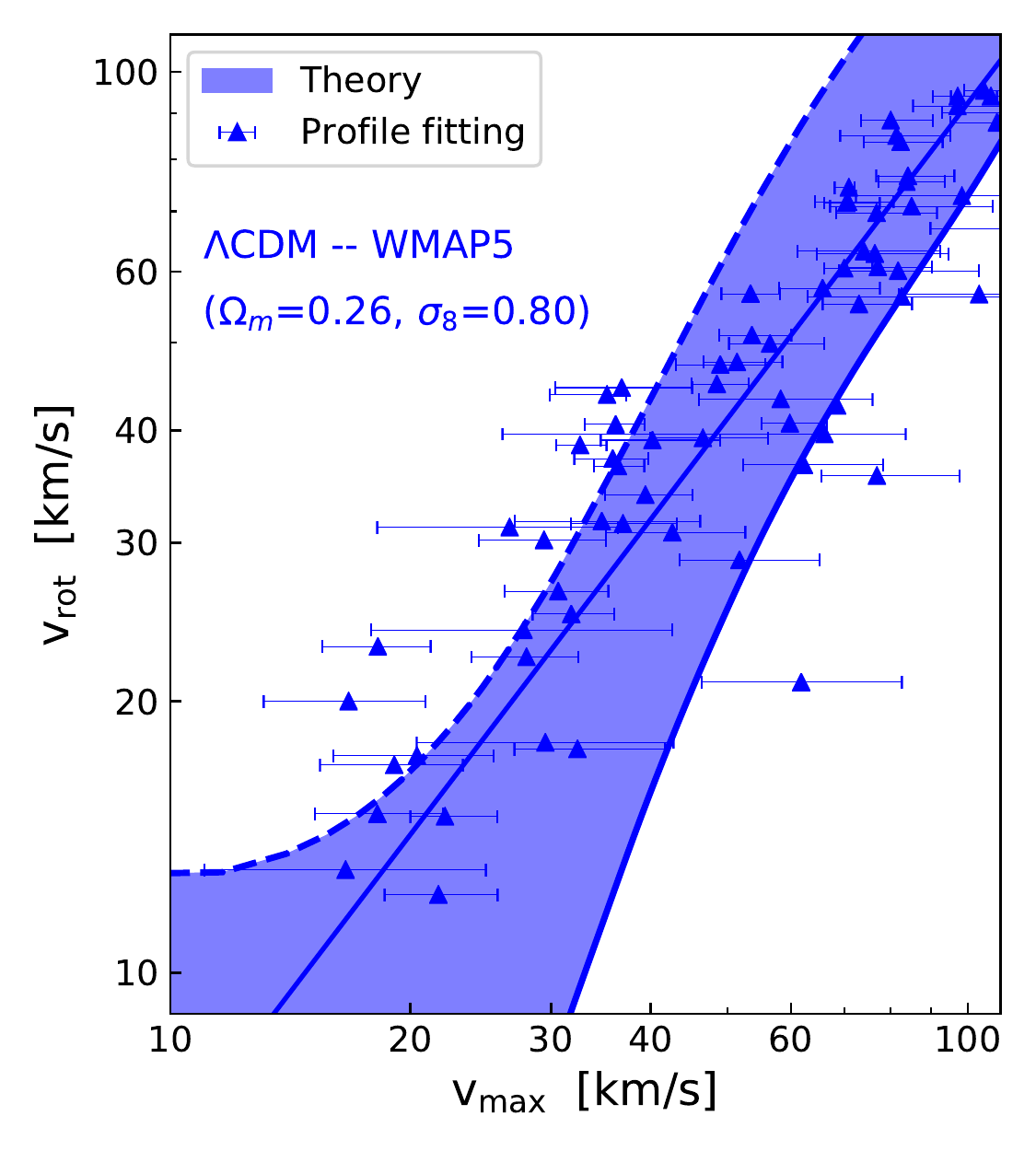}}
\adjustbox{trim={.03\width} {0.02\height} {0.02\width} {0.0\height},clip}{\includegraphics[width=.255\textwidth]{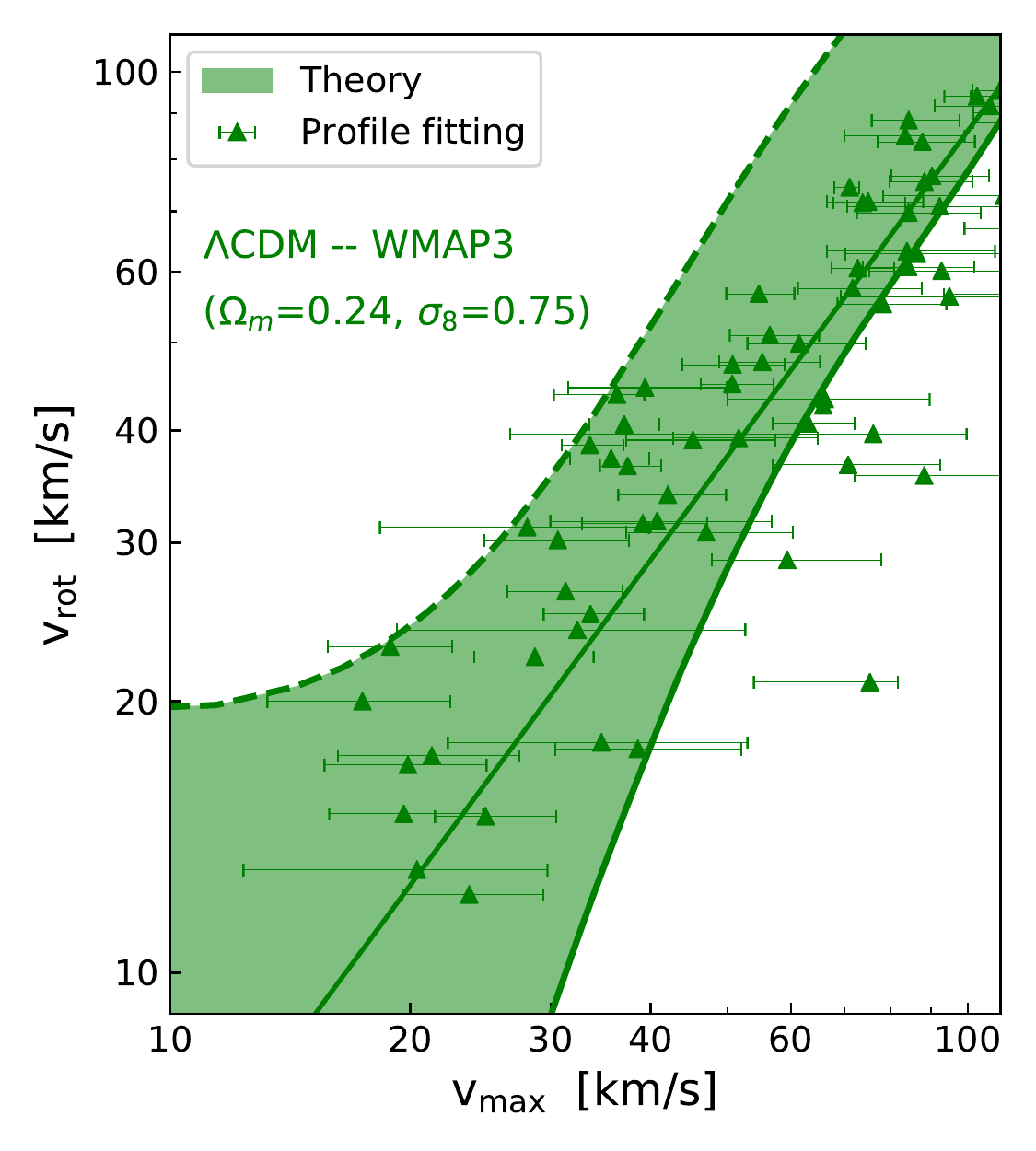}}
\caption{\label{fig:AMcomp}Relation between $v_{\rm rot}$ and $v_{\rm max}$ for dwarf galaxies with resolved kinematic measurements of the HI rotation (symbols with error bars) compared to the obtained relation from abundance matching the theoretical and observed galaxy velocity functions, including uncertainties from baryonic effects and sample variance (shaded areas). In the left panel we have added the results from \citep{Read:2016aaa} based on selected dwarf irregulars with high-quality rotation curves (magenta symbols, see text and caption of Fig.~\ref{fig:Planck} for more details). Both the profile fitting and the theoretical abundance matching lines are cosmology dependent (the former due to the concentration parameter). Colours refer to {\tt Planck} (black), {\tt WMAP7} (red), {\tt WMAP5} (blue), and {\tt WMAP3} (green). The rotation velocity ($v_{\rm rot}$) is defined via the HI line-width ($w_{50}$) and the galaxy inclination ($i$), i.e., $v_{\rm rot}\equiv w_{50}/(2\sin i)$.}}
\end{figure*}

\subsection{Relation between $v_{\rm rot}$ and $v_{\rm max}$ for {\tt Planck} and {\tt WMAP} cosmologies}\label{AMcomp}
The predicted velocity function based on reprojected HI line-widths ($v_{\rm rot}$) strongly depends on the connection between rotation and maximum circular velocities. The $v_{\rm rot}$-$v_{\rm max}$ relation, however, is based on the rotation-curve fitting of the selected galaxy sample which is a method with potentially important systematics and should only be trusted when averaging over many objects.

In this section we adopt a different perspective and compare theory and observations directly at the level of the $v_{\rm rot}$-$v_{\rm max}$ relation. In order to do this, the theoretical galaxy velocity function has to be related to observations by rank-ordering the mock halo sample using $v_{\rm max}$ and assigning a value of $v_{\rm rot}$ to each halo corresponding to the galaxy with the same rank in the Local Volume sample. In this way we obtain the $v_{\rm max}$-$v_{\rm rot}$ relation of DM haloes that is necessary to fit the observed velocity function. In principle, this description does not reveal more information than a direct comparison of the velocity functions, but it highlights the importance of the rotation curve fitting which is the most delicate step in our analysis.

In Fig.~\ref{fig:AMcomp} we plot the relation between $v_{\rm max}$ and $v_{\rm rot}$ of the selected sample of galaxies with spatially resolved kinematics (data points) and compare it to the relation needed to fit the observed galaxy velocity function assuming the {\tt Planck},  {\tt WMAP7}, {\tt WMAP5}, and {\tt WMAP3} cosmologies (from left to right). In all the cases, the average $v_{\rm rot}$-$v_{\rm max}$ relation obtained from profile fitting of kinematic data is well described by a linear regression line in logarithmic space (thin solid lines), and there is substantial scatter due to both observational errors and the inherent stochasticity of the concentrations. The regression line changes slightly when going from {\tt Planck} to {\tt WMAP} cosmologies, which is a direct consequence of the cosmology dependence of the concentration-mass relation (see Sec.~\ref{model}). In general, a decrease in $\Omega_{\rm m}$ or $\sigma_8$ leads to a decrease of the concentration and therefore an increase of the $v_{\rm max}$ estimate from fitting the HI kinematic data.

The theory predictions in Fig.~\ref{fig:AMcomp} are shown as broad colour-shaded bands exhibiting a rather strong cosmology dependence. They are enclosed by solid and dashed lines which correspond to the upper and lower limit of the velocity function (shown in Fig.~\ref{fig:VFcomp}) and also include the uncertainty due to sample variance of the observations.

For the {\tt Planck} cosmology, the predicted $v_{\rm rot}$-$v_{\rm max}$ relation is steeper than what the direct rotation curve estimates suggest\footnote{Recent papers have come to similar conclusions \citep{Papastergis:2014aba,Pace:2016oim,Papastergis:2016aba}, however, without including all potential effects from baryon physics.}. While the average relation from profile fitting (thin black line) agrees at large velocities, it lies outside of the predicted band below $v_{\rm max}\sim 45$ km/s. This is no longer true for the {\tt WMAP} cosmologies, where the kinematic fits reveal slightly larger values for $v_{\rm max}$ while at the same time the theory bands shift towards smaller velocities. The predicted $v_{\rm rot}$-$v_{\rm max}$ bands of all the {\tt WMAP} cosmologies overlap with the linear fits to the data (thin coloured lines) over the full velocity range.

The results shown in Fig.~\ref{fig:AMcomp} mean that in contrast to {\tt Planck}, predictions based on {\tt WMAP} cosmologies formally agree with observations when sample variance and maximum baryonic effects are included. However, a closer inspection reveals that for both {\tt Planck} and {\tt WMAP} cosmologies the predicted slope of the $v_{\rm rot}$-$v_{\rm max}$ relation seems to be slightly steeper than what a linear fit (in log-space) to the data points suggest. This could point towards an underestimated bias of the rotation-curve analysis, as already pointed out in Sec.~\ref{VFcomp}. However, the detailed analysis of the full rotation curves of a few dwarf irregulars with excellent data by Read17 yields a $v_{\rm rot}$-$v_{\rm max}$ slope in excellent agreement with our estimate, supporting the conclusion that a strong systematic bias in the fitting procedure is unlikely.

Note that Fig.~\ref{fig:VFcomp} and Fig.~\ref{fig:AMcomp} contain nearly the same information. However, Fig.~\ref{fig:AMcomp} specifically shows how crucially the cosmological constraints depend on the rotation-curve analysis, which is based on a limited amount of data. For example, the result for the {\tt Planck} cosmology are only discrepant at low velocities below $v_{\rm max}\sim45$ km/s, and there are only about 30 dwarf galaxies in this regime. This is a very low number if we consider that some of the galaxies might have important systematical errors. Indeed, it is possible that hidden biases related to observations \citep[resulting from erroneous inclination or distance measurements, see][]{Read:2016aaa} or the rotation-curve analysis \citep[related to the strong influence of stellar feedback on dwarf galaxy formation, see e.g. ][]{Maccio:2016egb,Verbeke:2017rfd} affect our results. In the near future, however, a wealth of new data from upcoming HI surveys will provide much better statistics, allowing to populate the small velocity regime with hundreds of new dwarf galaxies (see Sec.~\ref{prospects} for a detailed discussion).

\subsection{Cosmological parameter estimation}\label{WL}
In the previous sections we used the best-fitting cosmologies from {\tt Planck} and {\tt WMAP} to illustrate the sensitivity of the velocity function to changes in the cosmological parameters. Now we focus on the parameters $\Omega_m$ and $\sigma_8$ which determine the clustering amplitude at small scales and are therefore the parameters the velocity function is most sensitive to. In order to simplify the analysis, we keep all the other parameters of the $\Lambda$CDM model at the fiducial values of $\Omega_{\Lambda}=1-\Omega_m$, $\Omega_b=0.048$, $n_s=0.96$, and $H_0=68$, in agreement with CMB measurements. For the time being, we also set the neutrino masses to zero ($m_{\nu,1,2,3}=0$) but we will discuss the effects of nonzero neutrino masses in Sec.~\ref{neutrinos}. 

The goal of this section is to investigate for which values of $\Omega_m$ and $\sigma_8$ the predicted velocity function agrees with observations and for which values it either under- or over-predicts the abundance of galaxies. We define agreement by an overlap between theory (including all uncertainties due to baryonic effects) and observations (including uncertainties from sample variance) over the full range of velocities between $v_{\rm rot}=10$ km/s and $v_{\rm rot}=140$ km/s\footnote{We do not include higher values of $v_{\rm rot}$ because large galaxies have maximum circular velocities that are dominated by the stellar component. This means that it becomes much harder to connect $v_{\rm rot}$ to the halo mass (and therefore the cosmology). See Sec.~\ref{velbias} and Appendix~\ref{app:bias} for more details.}. Regarding Fig.~\ref{fig:VFcomp}, this means that the {\tt WMAP5} and {\tt WMAP3} cosmologies show agreement while the {\tt Planck} and {\tt WMAP7} cosmologies over-predict the observations.

In Fig.~\ref{fig:Oms8} we illustrate the parameter space of $\Omega_m$ and $\sigma_8$ where the area of agreement between predicted and observed velocity function is delimited by the dashed black line. The resulting region of parameter space is constrained form above and below because strong clustering (large $\sigma_8$ and $\Omega_m$) leads to an over-prediction of galaxy numbers at small scales ($v_{\rm rot}<50$ km/s), while weak clustering (small $\sigma_8$ and $\Omega_m$) results in an under-prediction at larger scales ($v_{\rm rot}>100$ km/s). The dashed black contour lines are obtained by calculating the theoretical velocity function for a grid of different $\sigma_8$ and $\Omega_m$ values, while keeping the other parameters at their fiducial value.

The region of agreement between the predicted and observed galaxy velocity function has significant overlap with the best fitting contour lines from the combined CMB experiments {\tt WMAP9+ACT+SPT}, the weak lensing survey {\tt KiDS} \citep{Hildebrandt:2016iqg}, as well as the X-ray cluster survey from \citet{Mantz:2014paa}. The same is true for other large-scale structure probes not shown in Fig.~\ref{fig:Oms8} such as the {\tt CFHTlens} weak lensing shear survey \citep{Heymans:2013fya}, the peak statistics of lensing \citep{Liu:2014wca,Kacprzak:2016vir}, or the Sunyaev-Zeldovich cluster counts \citep{Ade:2015fva}. The contours from {\tt Planck15} \citep{Planck:2015xua}, on the other hand, do not agree with the observed velocity function at the 2-$\sigma$ level. Our result can therefore be set into context with the previously reported tension of {\tt Planck} with other other cosmological probes \citep[e.g.][]{MacCrann:2014wfa,Nicola:2016qrc}.

Note that the findings presented in Fig.~\ref{fig:Oms8} are based on the concentration-mass relation from Eq.~(\ref{concentration}). This relation approximatively describes the cosmology-dependence of the concentrations and is constructed in a way that it gives the most accurate results for parameters close to the {\tt Planck} and {\tt WMAP5} cosmologies (see Appendix \ref{app:concentrations} for a test of the concentration-mass relation). Far away from this region of the parameter space, for example towards the top and bottom of the banana-shaped contour, the relation is expected to deteriorate, making the results less reliable.

\subsection{The role of massive neutrinos}\label{neutrinos}
It is well known that at least one of the three neutrino flavours must have a small but non-zero mass in the sub-eV range. In terms of structure formation this results in a smooth step-like suppression of the clustering signal at wave modes around $k\sim0.1$ h/Mpc. The exact shape of the suppression depends on the sum of the neutrino masses ($\Sigma m_{\nu}$) and the neutrino mass hierarchy. Since the velocity function probes significantly smaller scales beyond the step in the power spectrum, the value of $\Sigma m_{\nu}$ is expected to be highly degenerate with the primordial amplitude of scalar perturbations (often parametrised as $A_s$ or $\sigma_8$, respectively\footnote{For a cosmology with massless neutrinos, the amplitude of primordial perturbations ($A_s$) is directly proportional to $\sigma_8^2$ and they are interchangeably used as the fundamental parameter. In the case of massive neutrinos, on the other hand, this direct proportionality does not hold anymore. To avoid confusion we will use $A_s$ as the fundamental parameter for the amplitudes of perturbations in this section.}). The velocity function is therefore only expected to be useful for constraining the neutrino sector if the primordial amplitude of perturbations is measured by another survey that covers much larger scales.

In order to show how the velocity function is affected by the presence of neutrinos, we assume a cosmology with increasing $\Sigma m_{\nu}$ but otherwise constant reference parameters of $\Omega_{\rm m}$=0.3, $\Omega_{\rm b}$=0.048, $n_s$=0.96, $h$=0.68, and $A_s$=$2.335\times10^{-9}$. We choose this reference cosmology because it lies half-way between the best-fitting {\tt Planck} and weak lensing cosmologies.
In the top row of Fig.~\ref{fig:extensions}, we show the models with $\Sigma m_{\nu}=$ 0.0, 0.1 0.3, and 0.6 eV (from left to right). The case with massless neutrinos reveals a discrepancy between theory and observations, which is, however, smaller than for a cosmology with parameters set by {\tt Planck}. For massive neutrinos with $\Sigma m_{\nu}=0.1$ the discrepancy is further reduced but still visible, while for neutrino masses with $\Sigma m_{\nu}=0.3$ and above it disappears completely. 

\begin{figure}
\center{
\adjustbox{trim={.03\width} {0.06\height} {0.25\width} {0.09\height},clip}{\includegraphics[width=.68\textwidth]{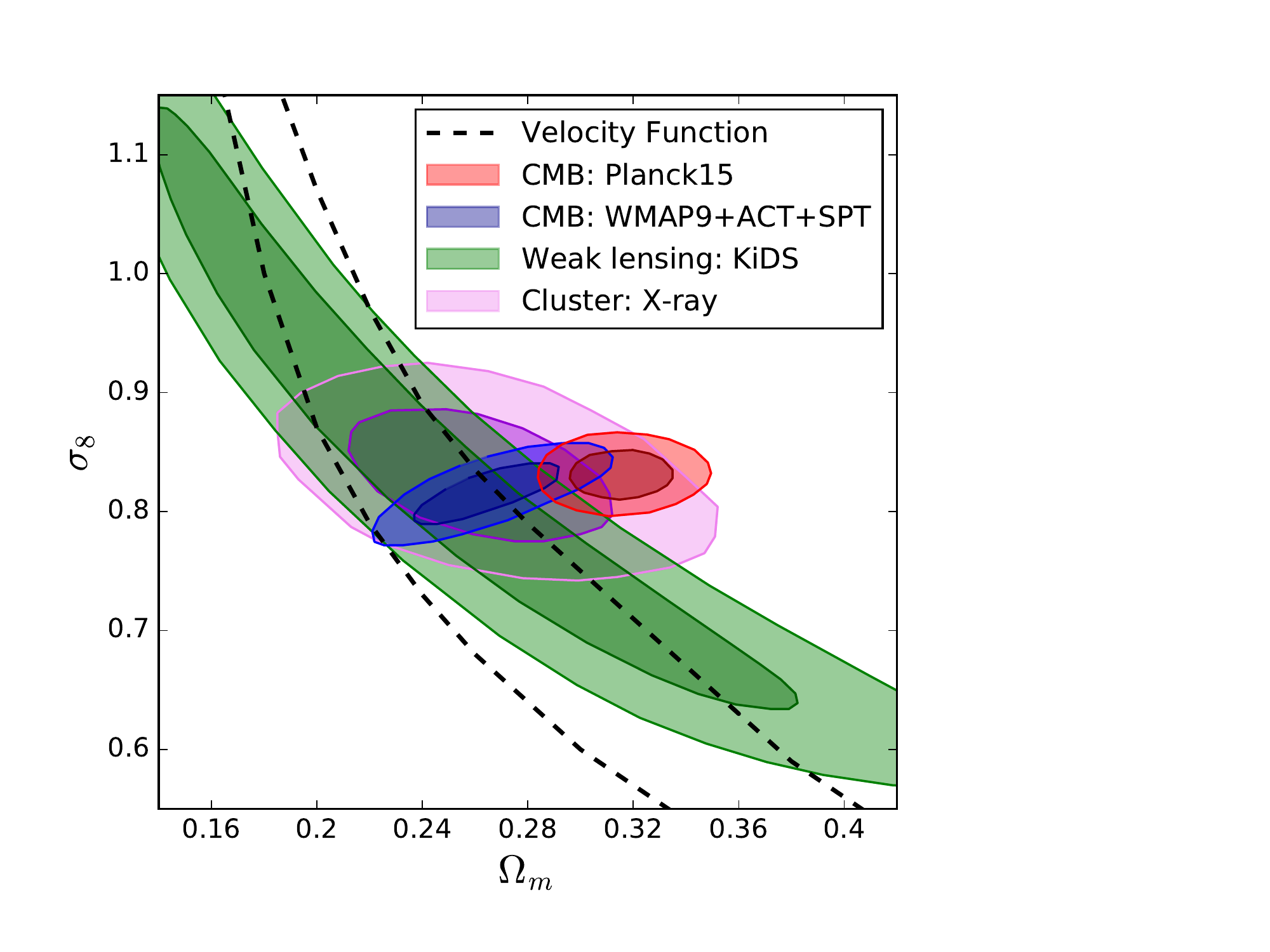}}
\caption{\label{fig:Oms8}Combined constraints on $\Omega_m$ and $\sigma_8$ from the velocity function of galaxies in the Local Volume (dashed black line) with all other cosmological parameters kept at the fiducial values (see text). The contours from the CMB measurements {\tt Planck} (red) and {\tt WMAP9+ACT+SPT} (blue) as well as the weak lensing survey {\tt KiDS} (green) and the cluster count observations of \citet[magenta]{Mantz:2014paa} are shown for comparison (bright and dark colours indicate 68\% and 95\% confidence levels)}.}
\end{figure}

Is it not surprising that increasing the neutrino mass in a cosmology with fixed scalar amplitude ($A_s$) reduces the predicted number density of galaxies as a function of rotation velocity. Indeed, larger neutrino masses lead to a stronger suppression of the small-scale clustering signal, which translates into a smaller value of $\sigma_8$ for a fixed $A_s$. The example cases shown in the top row of Fig.~\ref{fig:extensions} have values of $\sigma_8=$0.80, 0.77, 0.73, 0.66 (from left to right, the first one being the reference cosmology with massless neutrinos). From a look at Fig.~\ref{fig:Oms8}, it is clear that cosmologies with $\sigma_8$ between $\sim0.6$ and $\sim0.75$ should indeed yield a velocity function without discrepancy between theory and observations.

\begin{figure*}
\center{
\adjustbox{trim={.03\width} {0.02\height} {0.02\width} {0.0\height},clip}{\includegraphics[width=.255\textwidth]{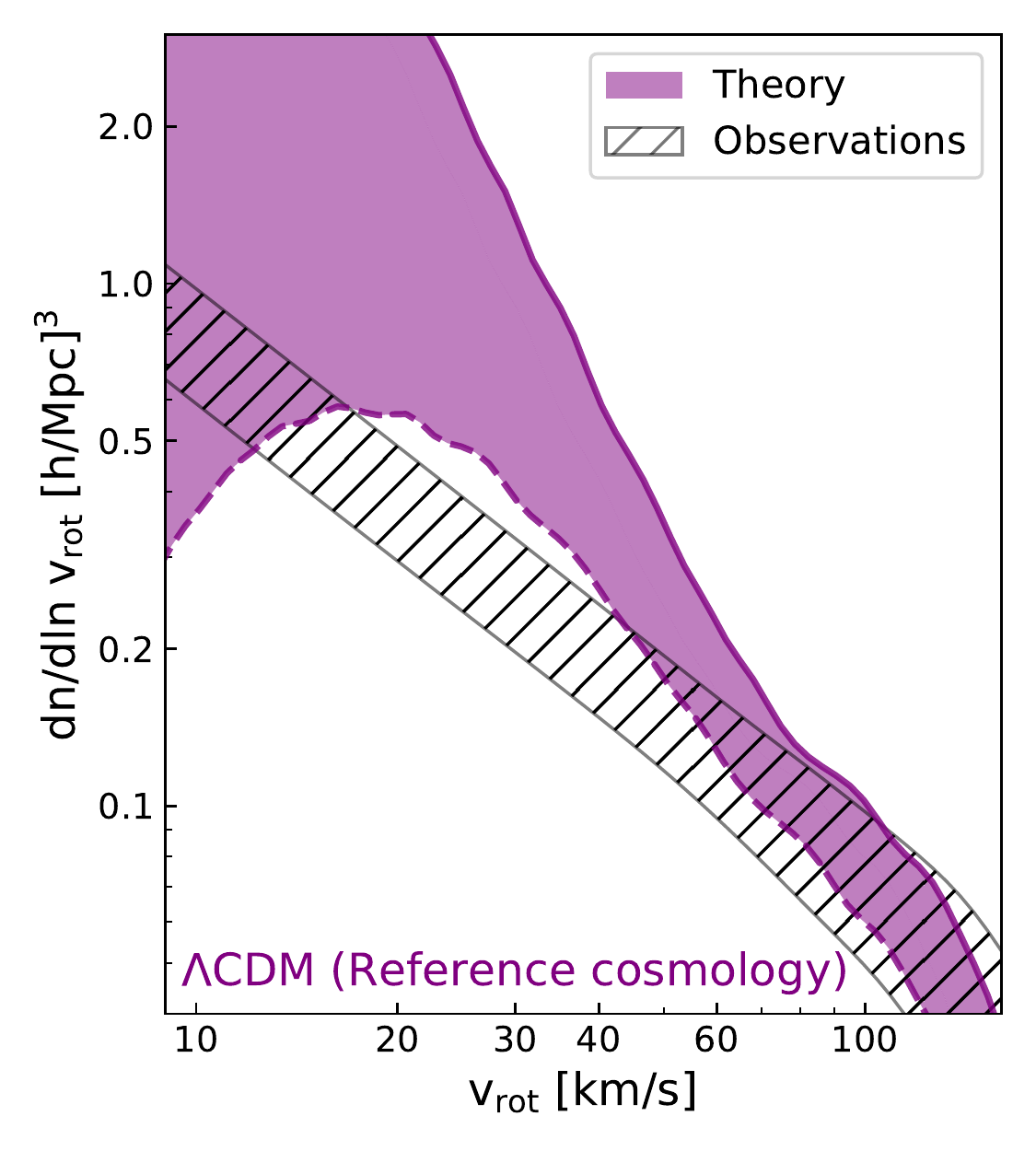}}
\adjustbox{trim={.03\width} {0.02\height} {0.02\width} {0.0\height},clip}{\includegraphics[width=.255\textwidth]{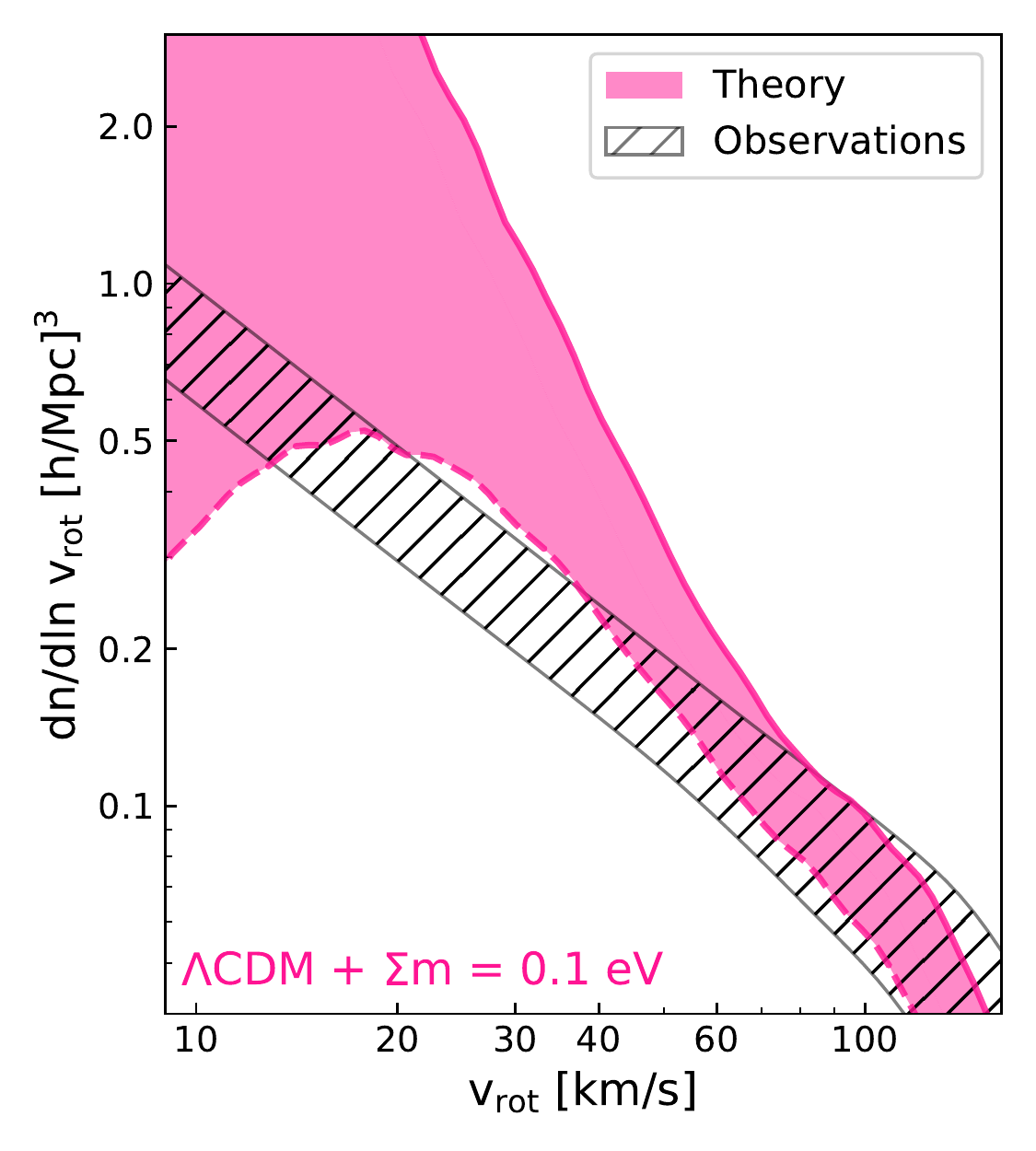}}
\adjustbox{trim={.03\width} {0.02\height} {0.02\width} {0.0\height},clip}{\includegraphics[width=.255\textwidth]{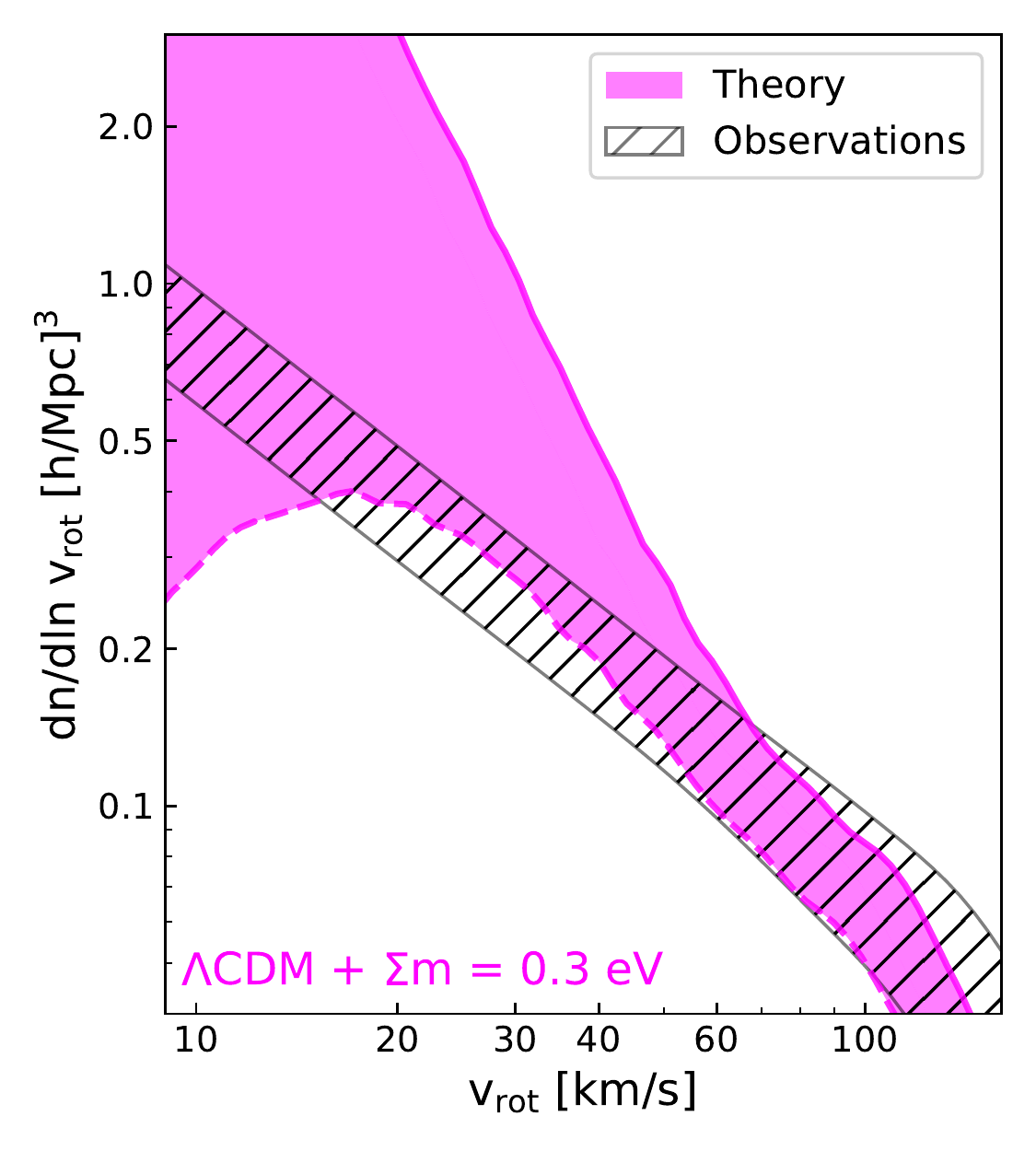}}
\adjustbox{trim={.03\width} {0.02\height} {0.02\width} {0.0\height},clip}{\includegraphics[width=.255\textwidth]{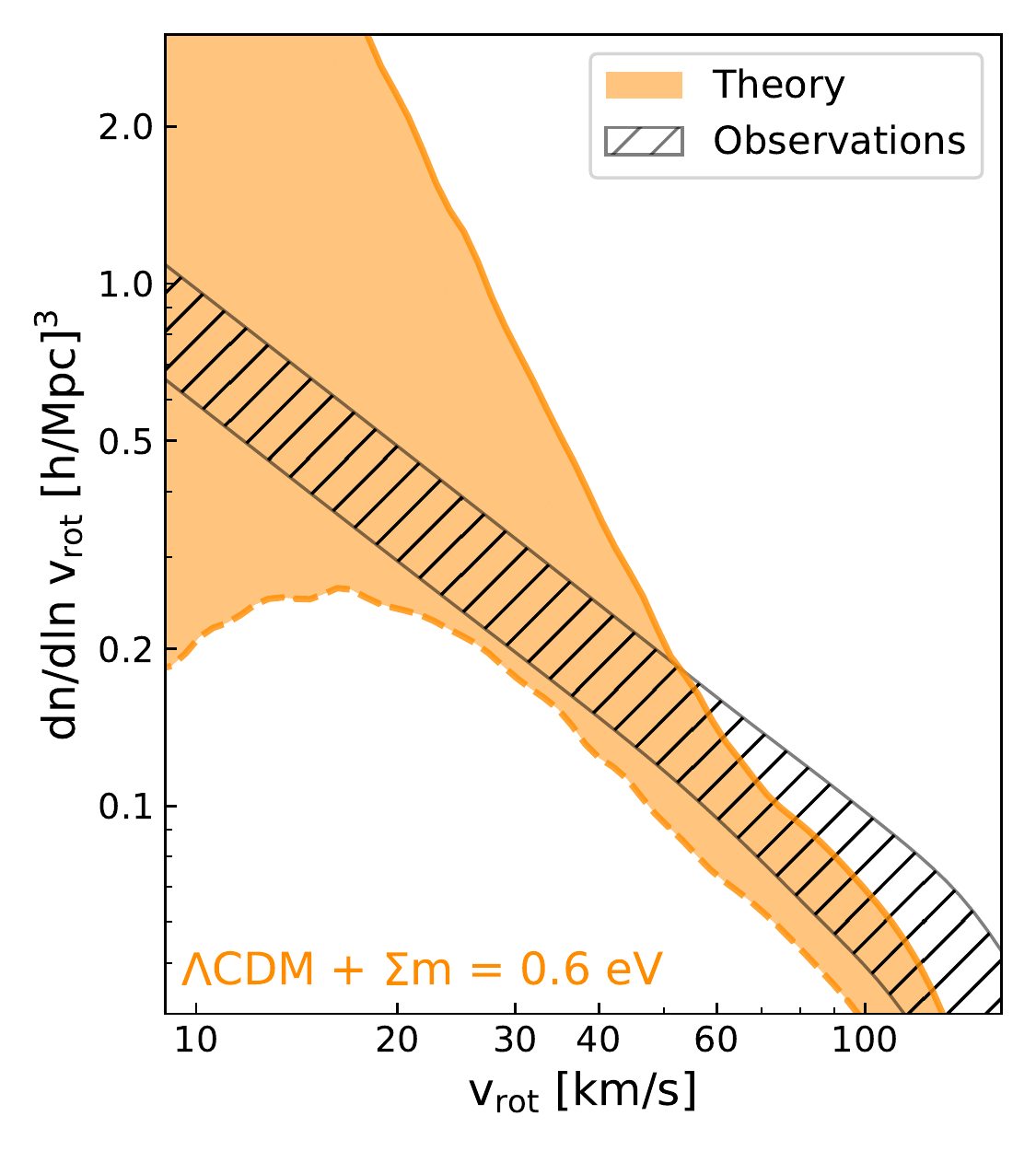}}\\
\adjustbox{trim={.03\width} {0.02\height} {0.02\width} {0.0\height},clip}{\includegraphics[width=.255\textwidth]{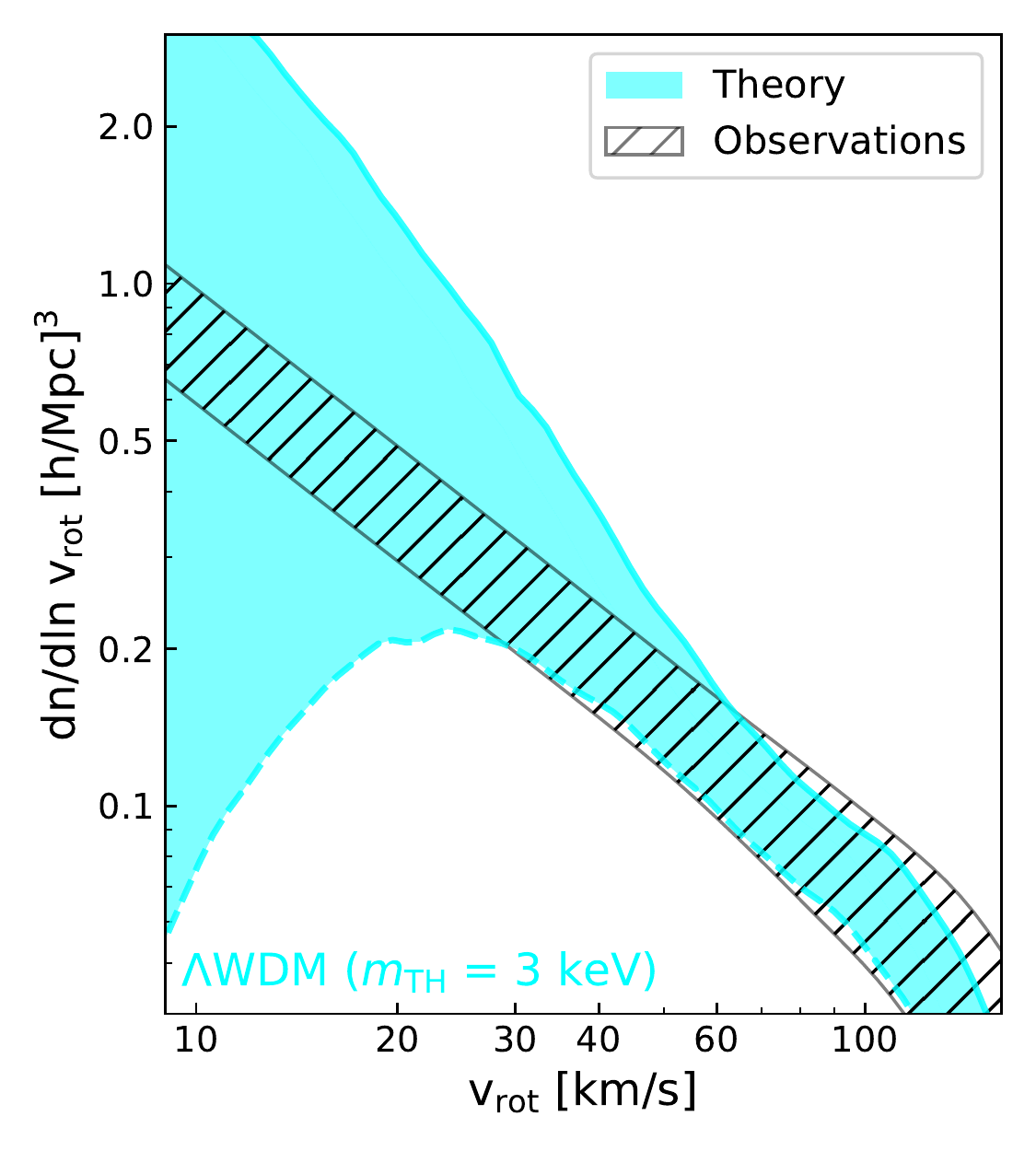}}
\adjustbox{trim={.03\width} {0.02\height} {0.02\width} {0.0\height},clip}{\includegraphics[width=.255\textwidth]{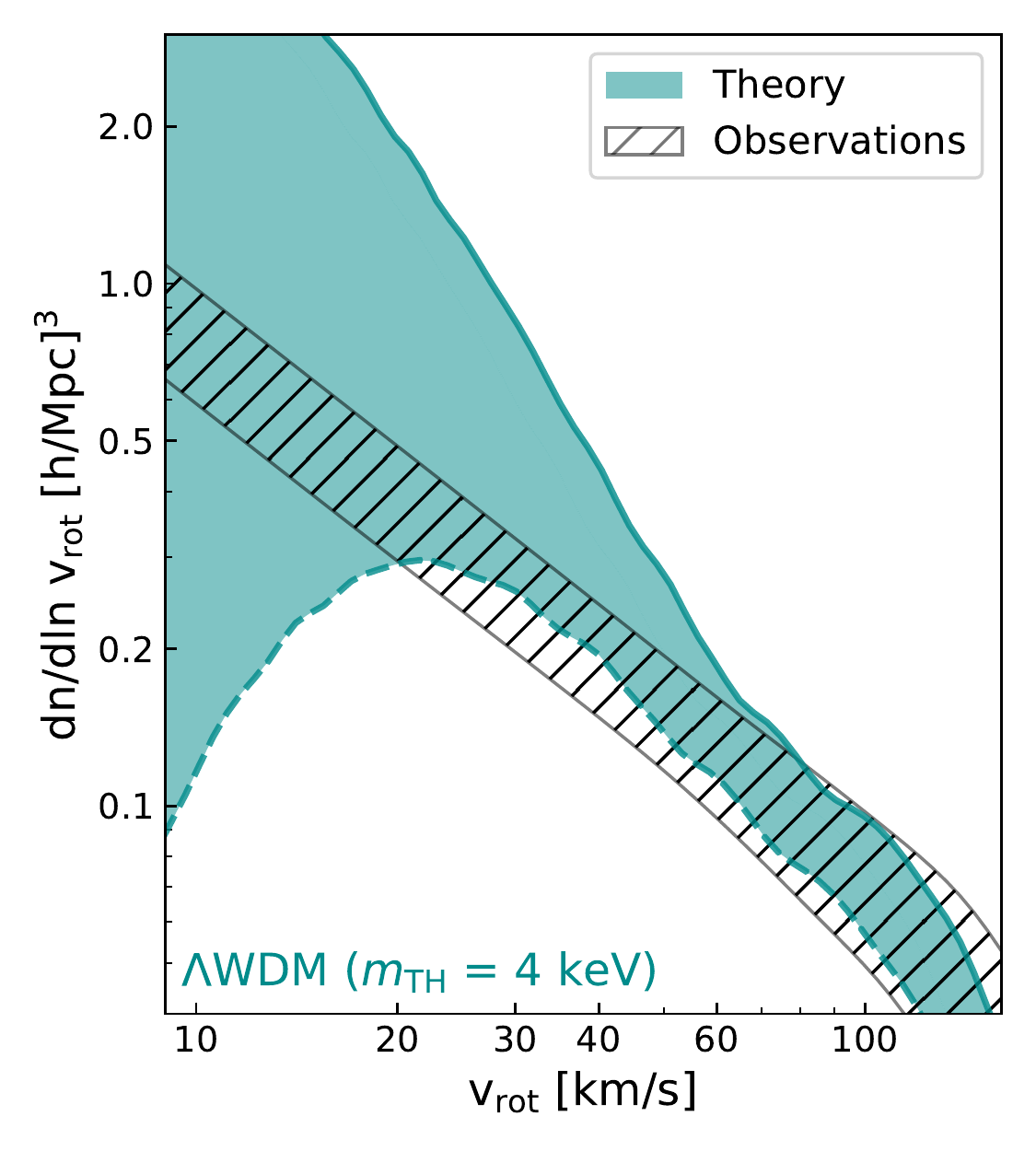}}
\adjustbox{trim={.03\width} {0.02\height} {0.02\width} {0.0\height},clip}{\includegraphics[width=.255\textwidth]{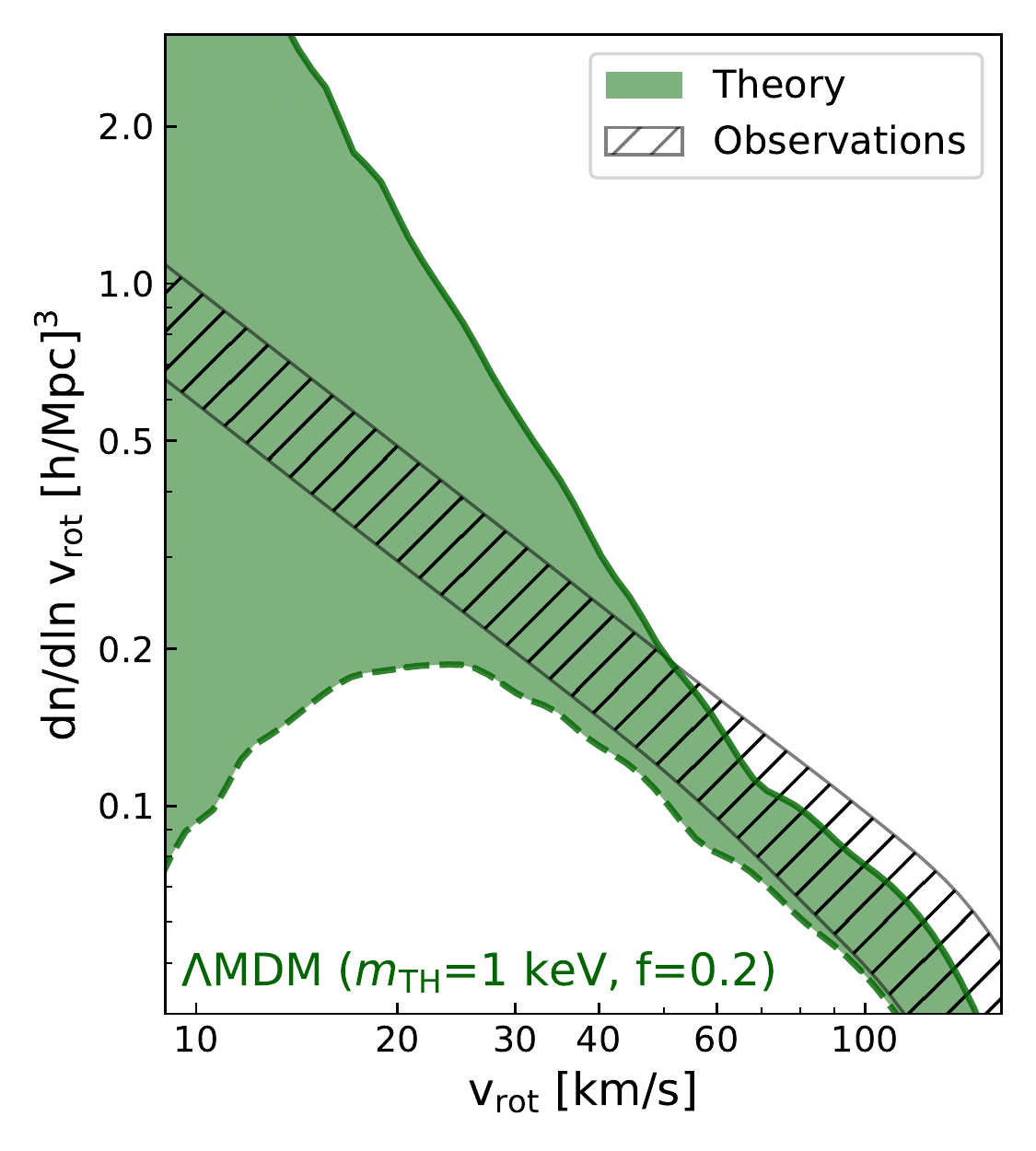}}
\adjustbox{trim={.03\width} {0.02\height} {0.02\width} {0.00\height},clip}{\includegraphics[width=.255\textwidth]{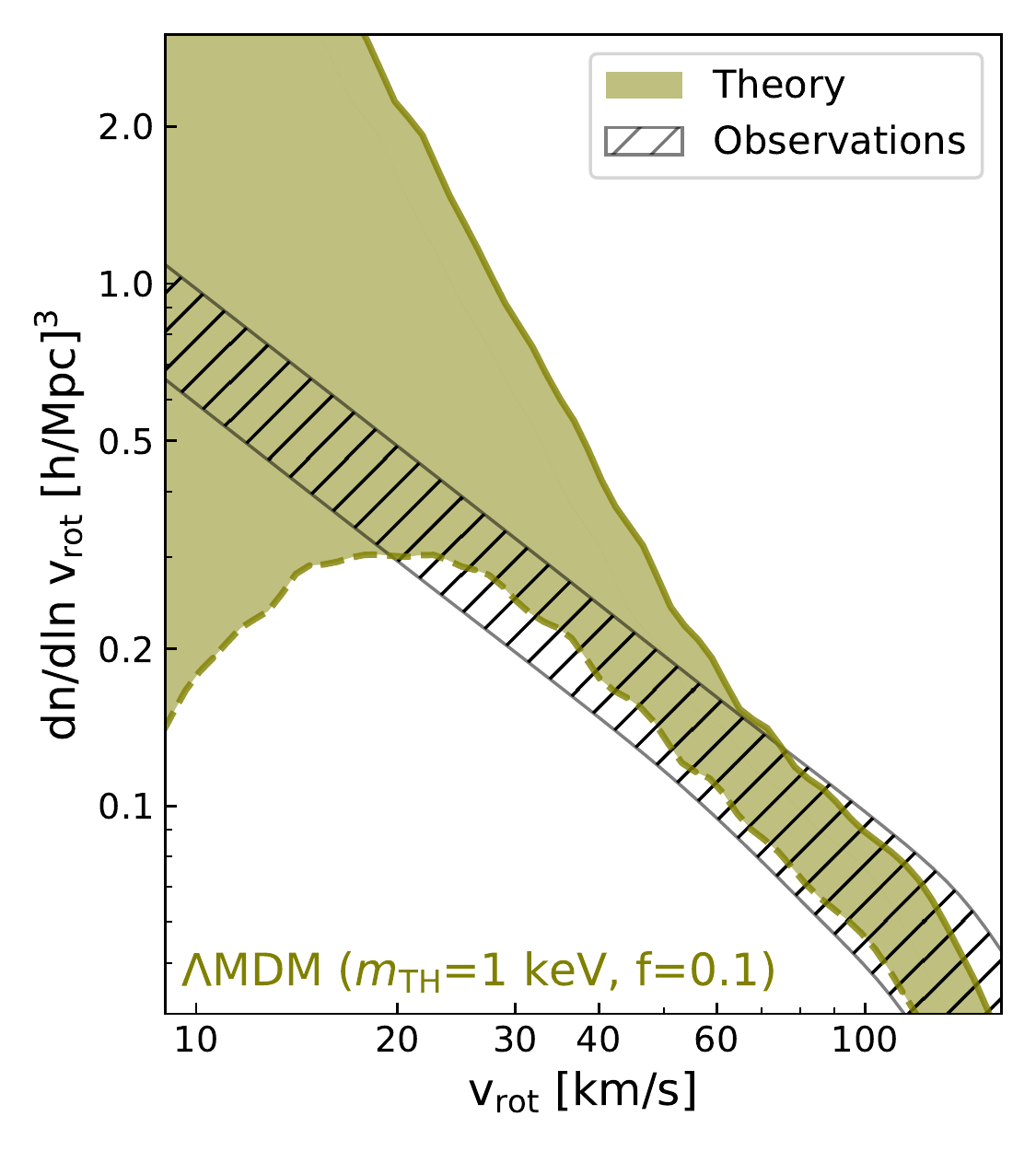}}
\caption{\label{fig:extensions}Theoretically predicted galaxy velocity function assuming the reference cosmology ($\Omega_{\rm}$=0.3, $\Omega_{\rm b}$=0.048, $n_s$=0.96, $h$=0.68, and $A_s$=$2.335\times10^{-9}$, consistent with both {\tt Planck} and large-scale structure probes) plus extensions of the neutrino and the dark matter sector. \emph{Top:} varying the sum of neutrino masses $\Sigma m$ = 0.0, 0.1, 0.3, 0.6 eV with normal mass hierarchy (from left to right). \emph{Bottom:} warm dark matter (with thermal relic mass $m_{\rm TH}$ = 3, 4 keV) and mixed dark matter (with WDM fraction $f$ = 0.2, 0.1 and thermal relic mass $m_{\rm TH}$ = 1 keV) scenarios (from left to right). The hatched band reproduces the observed galaxy velocity function of the Local Volume.}}
\end{figure*}

\subsection{Testing the dark matter sector}\label{altDM}
The velocity function allows not only to probe the neutrino sector but, more generally, it is sensitive to the physics of the dark matter particle. For example, it can be used to constrain the presence of a fourth neutrino species (i.e. a sterile neutrino) with mass in the keV range or below, since such a scenario leads to a characteristic suppression of the power spectrum at medium and small scales.

In a recent paper, we investigated various alternative dark matter scenarios (such as warm, mixed, and self-interacting DM) and how they affect the predicted galaxy velocity function \citep{Schneider:2016ayw}. Contrary to previous work \citep[e.g.][]{Zavala:2009ms,Papastergis:2011xe,Schneider:2013wwa} we included a complete treatment of the baryonic effects following the method of \citet{Trujillo-Gomez:2016pix}.

In this section we repeat the analysis of \citet{Schneider:2016ayw} investigating the effects of warm and mixed DM. These scenarios are generic enough to include many aspects of sterile neutrinos as a dark matter component \citep[see e.g.][]{Boyarsky:2008xj,Merle:2014xpa}. Compared to \citet{Schneider:2016ayw} we use a slightly modified analysis pipeline\footnote{We now assume a simple power-law fit to the $v_{\rm rot}$-$v_{\rm max}$ data for all DM scenarios, which is a good description as long as the DM model is in the lukewarm regime. Furthermore, we include the 3-$\sigma$ uncertainties of the fit into the analysis (see Sec.~\ref{vrotvmax}), leading to slightly less stringent limits on the predicted velocity function. Regarding the concentration-mass relation, on the other hand, we still apply the method presented in \citet{Schneider:2016ayw} as it provides more accurate results than Eq.~(\ref{concentration}) for the case of warm and mixed DM. Note that all these changes of the analysis pipeline only have small effects on the resulting velocity function.} and we assume the reference cosmology of Sec.~\ref{neutrinos} (with parameters $\Omega_{\rm}$=0.3, $\Omega_{\rm b}$=0.048, $n_s$=0.96, $h$=0.68, and $A_s$=$2.335\times10^{-9}$, and $m_{\nu,i}=0$, $i=1,2,3$) instead of the {\tt Planck} cosmology.

The lower row of Fig.~\ref{fig:extensions} shows the theoretical expectation for the galaxy velocity function for warm and mixed DM scenarios. Both warm DM models with a thermal relic mass of $m_{\rm TH}=3$ and 4 keV bring predictions and observations into better agreement compared to pure CDM. Note that these are reasonable DM models in agreement with conservative Lyman-$\alpha$ constraints \citep[see e.g.][]{Viel:2013apy,Garzilli:2015iwa,Irsic:2017ixq}. The mixed DM scenarios shown in Fig.~\ref{fig:extensions} assume a DM sector that is predominantly cold with a small fraction of $f=0.2$ and 0.1 consisting of a warm species with $m_{\rm TH}=1$ keV (where $f=\Omega_{\rm WDM}/(\Omega_{\rm WDM}+\Omega_{\rm CDM})$ and where $m_{\rm TH}$ is the thermal relic mass of the warm DM component). Such models with subdominant hot or warm DM components can alleviate potential tensions of the velocity function while passing stringent tests from Milky-Way satellite numbers \citep{Schneider:2014rda,Merle:2015vzu} and the Lyman-$\alpha$ forest \citep{Boyarsky:2008xj,Baur:2017stq}.

Compared to the results from \citet{Schneider:2016ayw}, even cooler dark matter scenarios (with $m_{\rm TH}\gtrsim4$ keV for warm and $f\lesssim0.1$ for mixed DM) are able to bring the predicted and observed galaxy velocity functions into agreement. This is because here we assume a different reference cosmology with slightly reduced $\Omega_{\rm m}$ and $\sigma_8$ compared to the {\tt Planck} cosmology. The differences between the results from this paper and from \citet{Schneider:2016ayw} show that the velocity function is very sensitive to both dark matter physics and cosmology, but there are partial degeneracies between the two sectors.

Next to warm and mixed DM, there are various other dark matter scenarios which either suppress small-scale perturbations or lead to modifications of inner density profiles and are therefore ideal candidates to be tested with the velocity function. Obvious candidates are ultra-light axion DM \citep{Marsh:2015xka,Hui:2016ltb}, interacting DM \citep{Lesgourgues:2015wza,Boehm:2017nrl}, self-interacting DM \citep[e.g.][]{Loeb:2010gj,Kamada:2016euw,Agrawal:2016quu}, decaying DM \citep{Enqvist:2015ara}, or sterile neutrinos from non-standard production mechanisms \citep{Adhikari:2016bei}.

\subsection{Other extensions of $\Lambda$CDM}\label{ext}
In principle, the velocity function has the potential to probe various extensions of the standard cosmological model which are not related to the dark matter sector. The most constraining power is thereby expected for all scenarios with altered small-scale clustering: first because these scales are directly tested by the velocity function and second because they are poorly constrained by probes from the CMB and large-scale structure statistics.

Probing the scale-dependence of the spectral index $n_s$ is an obvious target for investigation using the galaxy velocity function. While $n_s$ is assumed to be constant in the standard model of cosmology, some inflationary scenarios predict a positive or negative running of the spectral index. In terms of clustering, the effect from a running spectral index becomes strongest at the very large and very small observable scales, making the dwarf galaxy regime an ideal place to look for such an effect \citep[see, e.g.][]{Garrison-Kimmel:2014kia}. 

As a low-redshift and small-scale cosmological probe, the velocity function is well suited to test models of dark energy and modified gravity. For example, the dark energy equation of state (usually parametrised with $w_0$ and $w_a$) can be constrained efficiently with small-scale data, as any deviation from the standard model (i.e. $w_0=-1$ and $w_a=0$) has a strong effect on the non-linear clustering \citep[e.g.][]{McDonald:2005gz}. The same is true for modified gravity models which are subject to Chameleon or Vainshtein screening effects. For these models the strongest deviations from $\Lambda$CDM are observed at small clustering scales \citep[e.g.][]{Lombriser:2015axa}.


\section{Current limitations and future prospects}\label{prospects}
The results of this work are based on both the Local Volume galaxy catalogue \citep{Karachentsev:2013ipr} and a subsample of galaxies with spatially resolved HI kinematics \citep[see][]{Trujillo-Gomez:2016pix}. The former consists of about 900 galaxies most of them featuring $w_{50}$ measurements, the latter is a smaller sample with 109 galaxies, roughly one-third of them are galaxies with $v_{\rm rot}<50$ km/s. These are rather small numbers for a statistical analysis, making the current results susceptible to systematic effects. In the following, we discuss the most important potential systematics, before giving an outlook on the expected observational progress and how this progress will improve the viability of the velocity function as a cosmological probe.

The analysis of this paper assumes the selected galaxies to be a representative sample of the all galaxies in the Local Volume. In \citet{Trujillo-Gomez:2016pix} we performed several test and found both samples to be consistent, but note that excluding hidden biases is a difficult task. For example, the selected galaxies were chosen to have kinematic measurements at radii well outside of the stellar disk, ensuring that the measurement is unaffected by the potential presence of a core induced by feedback effects. This choice is only viable if the ratio between gas and stellar disk sizes is independent of halo characteristics, which is a reasonable assumption but currently untested. We know, however, that the selection criterion does not affect any observed galaxy properties \citep[like for example the slope of the average $M_{\rm b}$-$v_{\rm rot}$ relation, see][]{Trujillo-Gomez:2016pix}.

Another potential issue are erroneous estimates for the observed galaxy distances and inclinations \citep[see e.g.][]{Read:2016bbb}. However, while such observational errors are likely to affect some individual galaxies, they are unlikely to significantly bias the result \citep[][]{Papastergis:2016aaa}.

Arguably the most delicate part of the analysis in this paper is the estimate of $v_{\rm max}$ based on HI kinematics. As it is customary for a rotation curve analysis, we assume galaxies to have axisymmetric gas distributions in approximate equilibrium. The gravitational potential can then be obtained from the gas rotation speed and dispersion, assuming radial motions to be negligible. However, recent hydrodynamical simulations of dwarf galaxies have revealed that outbursts due to feedback may result in velocity fields that are temporarily out of equilibrium \citep{Teyssier:2012ie}, leading to an underestimation of the maximum circular velocity \citep{Verbeke:2017rfd}\footnote{Mock observations based on other simulations, on the other hand, find an overestimation of $v_{\rm max}$ instead \citep{Oman:2017vkl}.}. In principle it is possible to only select dwarf galaxies in approximate equilibrium but this would require a larger sample size.

Finally, it is possible that the galaxy sample of the Local Volume is incomplete. Such a scenario would, however, require a large number of undetected very low-surface brightness galaxies which K15 has argued to be highly unlikely.

In the near future the ongoing large area HI surveys {\tt WALLABY} and {\tt WNSHS} (both pathfinders of the ultimate {\tt SKA} survey) will detect a large number of new dwarf galaxies, a notable fraction of them with sufficient resolution to allow for high-quality measurements of gas kinematics. {\tt WNSHS} is expected to find $\sim5\times10^4$ HI sources in an area of 3500 deg$^2$ of the northern sky with a resolution of $15^{\prime\prime}$ \citep{Giovanelli:2015aaa}. This means that about 30000 galaxies are expected to have at least one kinematic measurement, i.e. they will extend across 3 beams or more. Approximatively 1000 of those galaxies will have well resolved rotation curves, with the observable HI extend spreading across ten or more beams. In terms of dwarf galaxies with rotation velocities in the critical regime below $v_{\rm rot}\sim 50$ km/s, we expect about 800 objects to have one kinematic measurement and 20-30 to feature well resolved rotation curves.

{\tt WALLABY} will be directed towards the southern sky and detect at least $5\times10^5$ HI sources in an area of 30000 deg$^2$ \citep{Koribalski:2016aaa}. Since {\tt WALLABY} has lower resolution ($30^{\prime\prime}$) but observes a much larger area than {\tt WNSHS}, it is expected to obtain rotation curves for a similar number of galaxies. Roughly 40000 galaxies are expected to be resolved with 3 beams or more, allowing for at least one kinematic measurement. About 1000 of these galaxies will feature well resolved rotation curves with their HI component extending over ten or more beams. At $v_{\rm rot}\sim 50$ km/s or lower, we expect about 1000 galaxies with at least one kinematic measurement and more than 30 with well resolved rotation curves. 

Within the next two years, {\tt WNSHS} and {\tt WALLABY} will bring dynamical modelling studies to a new level. Compared to the currently available data they will increase the number of dwarf galaxies with kinematic information by more than a factor of 50. Some of these galaxies will feature very well resolved rotation curves that can be used for detailed analysis of the underlying potential. 

The upcoming wealth of new data will allow us to obtain a much more precise relation between $v_{\rm rot}$ and $v_{\rm max}$ based on blindly selected galaxies. We will be able to filter out potentially biased galaxies more efficiently, for example if they show signs of disequilibrium or if they have poorly known inclination. Even more importantly, the full data from {\tt WNSHS} and {\tt WALLABY} will make it possible to construct the velocity function directly from kinematic measurements (such as $v_{\rm out}$ at $r_{\rm out}$) without relying on unresolved HI line-widths which are known to be biased at small scales.


\section{Conclusions}\label{conclusions}
The number density of galaxies as a function of rotation velocity -- the galaxy velocity function -- is a powerful statistical measure that probes the very small scales of cosmological structure formation. The main advantage of the velocity function compared to other probes is that it can be directly connected to theory since the observed rotation velocities of galaxies trace the underlying halo (dark matter and baryon) potential. This characteristic distinguishes the velocity function from galaxy counts with respect to stellar or gas mass, for example, which depend sensitively on poorly known details of the physics of galaxy formation and are therefore much harder to connect to theory.

In this paper we investigate for the first time the potential of the galaxy velocity function as a cosmological probe. We find that changes in cosmological parameters such as the amplitude of perturbations ($\sigma_8$) or the total matter abundance ($\Omega_{\rm m}$) can have surprisingly strong effects on the velocity function. This is because the signal depends on a combination of halo abundance (which produces vertical shifts in the velocity function) and concentrations (which shifts the velocity function horizontally by changing the rotation velocity for a fixed halo mass), both effects are cosmology dependent and complement each other. 

In an earlier series of two papers, we developed a framework to factor in all potential effects from baryonic physics that are relevant for the predictions of the theoretical velocity function \citep{Trujillo-Gomez:2016pix, Schneider:2016ayw}. The main conclusion of this work was that even when including maximum suppression effects from baryons, the theory prediction based on the {\tt Planck} cosmology cannot be brought into agreement with observations. In the present paper, we show that this discrepancy can be strongly alleviated when assuming best-fitting cosmologies from weak lensing surveys or galaxy cluster counts, instead. It is intriguing that the velocity function agrees with other large-scale structure surveys but not with the latest CMB measurements.

In addition to varying standard cosmological parameters, we also investigated the effects of massive neutrinos on the amplitude of the velocity function. While increasing the sum of the neutrino masses ($\Sigma m_{\nu}$) helps to alleviate the initial discrepancy, the effect is perfectly degenerate with varying the clustering amplitude ($\sigma_8$). This is not surprising as the velocity function probes small scales well beyond the steplike suppressions of the clustering signal induced by the neutrinos.

Finally, we also studied how warm or mixed dark matter (DM) scenarios affect the galaxy velocity function. When assuming cosmological parameters in (2-$\sigma$) agreement with both the CMB survey {\tt Planck} and the weak lensing survey {\tt KiDS}, we find that \emph{lukewarm} models with thermal mass of $m_{\rm TH}\gtrsim4$ keV (for warm DM) and $f\lesssim0.1$ (for mixed DM, where $f$ is the fraction of warm to total DM abundance) are able to bring the predicted velocity function into agreement with the observations. These scenarios are even \emph{cooler} than the ones found in our previous work based on the {\tt Planck} reference cosmology \citep{Schneider:2016ayw}, easily avoiding constraints from the Lyman-$\alpha$ forest.

Our current analysis of the velocity function is based on the kinematic information of about one hundred galaxies, one third of them being in the most important regime of dwarf galaxies. This is a small number, especially when considering that some of the measurements could be biased due to erroneous inclinations and distance measurements or due to oversimplifications in the rotation-curve analysis \citep[see e.g.][]{Read:2016bbb,Verbeke:2017rfd}. The current result should therefore be taken with a grain of salt.
However, upcoming data from the currently ongoing wide-field HI surveys {\tt WNSHS} and {\tt WALLABY} will be a game-changer for dynamical studies of galaxies. At dwarf galaxy scales, there will be about a factor of 50 more galaxies with kinematic information than what is currently available. Next to dramatically reducing statistical uncertainties, this will allow us to simplify the analysis and circumvent potential systematics by constructing the velocity function directly from kinematical observations. With the new data at hand, we are confident that the velocity function will become a reliable cosmological probe at scales well below the reach of galaxy clustering or weak lensing surveys.


\section*{Acknowledgements}
We thank Emmmanouil Papastergis and Andrina Nicola for their help. AS acknowledges support from the Swiss National Science Foundation (PZ00P2\_161363).


\bibliographystyle{mnras}
\bibliography{ASbib}


\appendix
\section{Testing the extended Press-Schechter model of the velocity function}\label{app:EPStest}
In this paper we use the extended Press-Schechter (EPS) approach with a sharp-$k$ filter \citep[developed in][]{Schneider:2013wwa,Schneider:2016ayw} to predict the velocity function for different cosmologies. The EPS method consists of an analytical calculation of non-linear structure formation providing approximate results. It is therefore important to check if these results are accurate enough for the purpose of the present analysis.

Here we compare the EPS velocity function to the results from the {\tt Bolshoi} \citep{Klypin:2010qw} and {\tt BolshoiP} \citep{Klypin:2014kpa} $N$-body simulations. These simulations are based on cosmological parameters close to the values from {\tt WMAP7} and {\tt Planck}, and they include precise measurements of the concentrations-mass relation \citep[see][]{Klypin:2014kpa}.

\begin{figure}
\center{
\adjustbox{trim={.03\width} {0.06\height} {0.25\width} {0.09\height},clip}{\includegraphics[width=.68\textwidth]{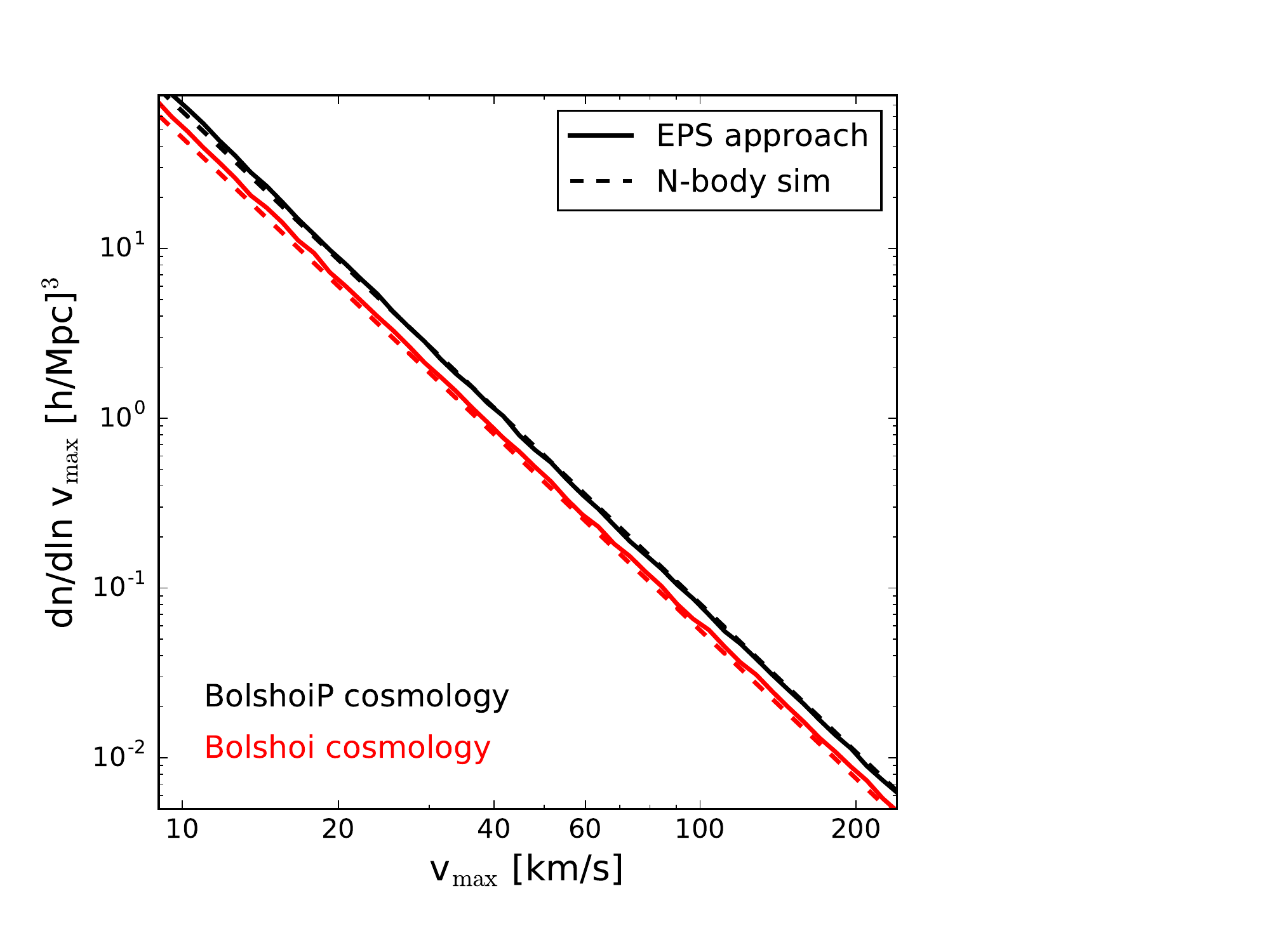}}
\caption{\label{fig:EPStest}Halo velocity functions based on the extended Press Schechter (EPS) approach (solid lines) compared to results from the {\tt Bolshoi} \citep{Klypin:2010qw} and {\tt BolshoiP} \citep{Klypin:2014kpa} $N$-body simulations (dashed lines). Note that these simulations are based on cosmological parameters close to {\tt WMAP7} and {\tt Planck}, respectively.}}
\end{figure}

In Fig.~\ref{fig:EPStest} we plot the DM halo velocity function as a function of $v_{\rm max}$ obtained with the EPS model (and without including baryonic effects) assuming {\tt Bolshoi} (red solid line) and {\tt BolshoiP} (black solid line) cosmologies. The EPS results are in excellent agreement with the velocity functions directly measured from the simulations (red and black dashed lines for {\tt Bolshoi} and {\tt BolshoiP}, respectively). This confirm the validity of our model to provide accurate predictions for the galactic velocity function. Note, however, that this is only true if the concentration-mass relation is known well enough. In Appendix \ref{app:concentrations} we further discuss the accuracy of the concentration-mass relation used in this paper.

\section{Testing the concentration-mass relation}\label{app:concentrations}
The DM halo concentrations are a crucial ingredient for predicting the velocity function using an extended Press Schechter approach. Unfortunately, there are currently no complete models for the concentration-mass relation which are applicable to different cosmologies, cover scales down to the dwarf-galaxy regime, and are sufficiently accurate for the purpose of this paper. The only existing cosmic emulator for concentrations based on $N$-body simulations \citep{Kwan:2012nd} does not have the resolution to include objects below the Milky-Way mass scale.

In this paper, we use the model from \citet{Diemer:2014gba} which extends to the scales of dwarf galaxies and is designed to cover arbitrary cosmologies within the $\Lambda$CDM framework. Since it provides only approximate results, we used the model to predict the relative changes of the concentration-mass relation with respect to the {\tt Planck} cosmology (see Eq. \ref{concentration} in Sec. \ref{model}). This guarantees very accurate results for cosmologies with parameters that do not deviate too much from the {\tt Planck} values.

\begin{figure}
\center{
\adjustbox{trim={.03\width} {0.06\height} {0.25\width} {0.09\height},clip}{\includegraphics[width=.68\textwidth]{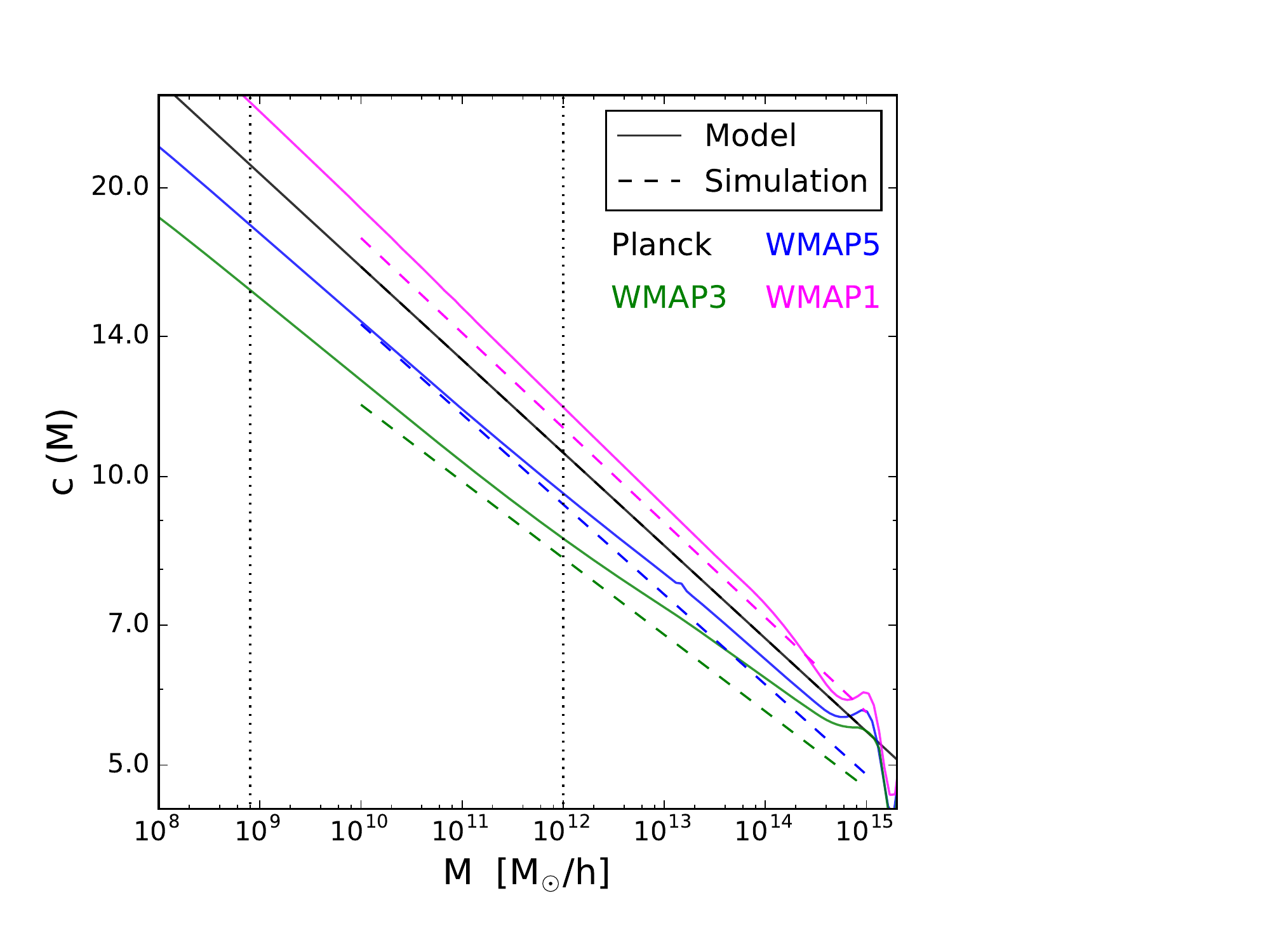}}
\caption{\label{fig:concentration}Concentration-mass relation from Eq.~(\ref{concentration}) compared to direct measurements from simulations with different cosmologies \citep{Maccio:2008pcd,Dutton:2014xda}. The relevant scales for the velocity function are delimited by vertical dotted lines. By construction, the agreement between model and simulations is best for cosmologies close to {\tt Planck} best fitting parameters.}}
\end{figure}

In Fig. \ref{fig:concentration} we compare the concentration-mass relation obtained from Eq. (\ref{concentration}) with direct fits to simulations based on {\tt Planck}, {\tt WMAP1}, {\tt WMAP3}, and {\tt WMAP5} cosmologies \citep{Maccio:2008pcd,Dutton:2014xda}. The dotted vertical lines delimit the mass range relevant for the velocity function. Our model extends below masses of $10^{10}$ M$_{\odot}$/h, which is the resolution limit of the simulations. By construction, there is perfect agreement for the {\tt Planck} cosmology. For the case of {\tt WMAP5} the agreement is still very good, while the model starts to deviate somewhat from the simulation results for the {\tt WMAP3} and {\tt WMAP1} cosmologies.

The results from Fig. \ref{fig:concentration} show that the concentrations we use in this paper are accurate for the region of parameter space that includes best fitting cosmologies from CMB and weak lensing experiments. However, for other parts of the parameter space, for example towards the top and bottom of the banana-shaped contour in Fig. \ref{fig:Oms8}, the concentration-mass relation is expected to become less accurate.

\section{Effects of baryons on the \boldmath{$v_{\rm max}$} of massive galaxies}\label{app:bias}
Dwarf galaxies are highly dark matter dominated at all radii. As a consequence, the stellar an gas components never contribute substantially to the circular velocity profiles and $v_{\rm max}$ is always set by the dark matter component. This is not necessarily true as the mass increases towards Milky Way scales and beyond, where the massive and dense stellar component in the halo centre can dominate the velocity profile and lead to contraction of the dark matter. Both effects may increase the value of $v_{\rm max}$ of the galaxy compared to its host DM halo as obtained from $N$-body simulations or the EPS approach.

\begin{figure}
\center{
\adjustbox{trim={.03\width} {0.06\height} {0.25\width} {0.09\height},clip}{\includegraphics[width=.68\textwidth]{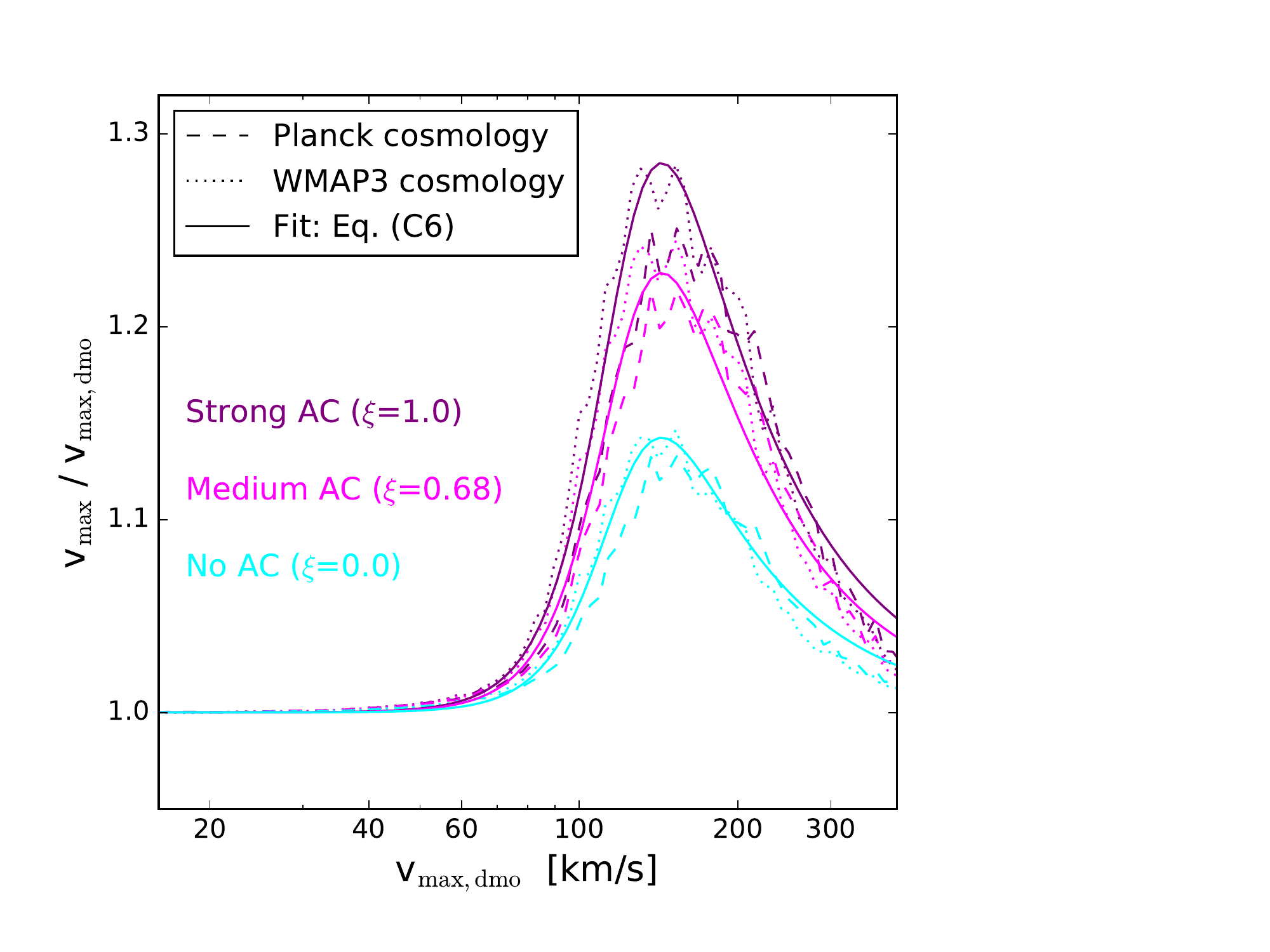}}
\caption{\label{fig:velbias}Boost of $v_{\rm max}$ due to the presence of a stellar disk, assuming strong (purple), medium (magenta), and no (cyan) adiabatic contraction (AC). Dashed and dotted lines assume {\tt Planck} and {\tt WMAP3} cosmologies, respectively, while solid lines correspond to the fitting function of Eq.~(\ref{fitvelbias}) with $v_c=125$ km/s and $\alpha=0.5,0.4,0.25$. The maximum circular velocity from gravity-only $N$-body simulations (or the EPS approach) is denoted by $v_{\rm max,\,dmo}$.}}
\end{figure}

We now attempt to model this velocity effect by adding the contributions of each mass component
\begin{equation}
v(r) = \sqrt{v_{\rm dm}^2(r)+v_{\rm gas}^2(r) + v_{\rm star}^2(r)}\,,
\end{equation}
where $v_{\rm star}$, $v_{\rm gas}$, and $v_{\rm dm}$ are the individual velocity profiles of stars, cold gas, and dark matter.

For the stellar component we assume an exponential disc \citep[e.g.][]{Binney2008}, i.e.
\begin{equation}
v_{\rm star}^2(r) = \frac{2GM_{\rm star}}{r_h}y^2\left[I_{0}(y)K_0(y)-I_1(y)K_1(y)\right]
\end{equation}
where $y=r/(2r_h)$ and where $I_0$, $I_1$, $K_0$, $K_1$ are modified Bessel functions of first and second kind. Based on the correlation observed between the half-light radius ($r_h$) and the stellar mass ($M_{\rm star}$) of galaxies in SDSS, we set 
\begin{equation}\label{size-mass}
r_h(M_{\rm star}) = r_0\left(\frac{M_{\rm star}}{M_0}\right)^{\alpha}\left[\frac{1}{2}+\frac{1}{2}\left(\frac{M_{\rm star}}{M_0}\right)^{\gamma}\right]^{(\beta-\alpha)/\gamma}
\end{equation}
with $\alpha=0.18$, $\beta=0.52$, $\gamma=1.8$, $M_0=2.75\times10^{10}$ M$_{\odot}$, and $r_0=5.25$ kpc \citep{Dutton:2010pz}. We furthermore assume a scatter  of $\sigma_{\ln r_h}\sim0.2$ at a fixed value of $M_{\rm star}$ \citep[see Fig.~1 of][]{Dutton:2010pz}. The stellar-to-halo mass relation is obtained using the abundance matching approach discussed in Sec.~\ref{AMcomp}. 
In addition to a stellar disk, bulges become increasingly common for stellar masses above $10^9$ M$_{\odot}$/h. However, we have checked that an additional bulge (with mass fraction of 0.33 and modelled by a Sersic profile with $n=0.25$ and scale radius $r_s=0.44r_h$) does not change our results, and we therefore ignore this additional correction for simplicity.

The cold gas component ($v_{\rm gas}$) is generally assumed to follow an extended exponential profile. However, given the observed subdominant gas fractions and the rather large scale lengths of the gas disk, we find that this component does not affect the value of $v_{\rm max}$. We therefore do not model this component, implicitly assuming the gas to follow the DM profile.

Finally, the DM component ($v_{\rm dm}$) can be modelled with a NFW profile, i.e.
\begin{equation}
v_{\rm dm}^2(r)= \frac{GM}{r}\left[\frac{\ln(1+cx)-cx/(1+cx)}{\ln(1+c)-c/(1+c)}\right],
\end{equation}
where $M$ is the halo mass, $c$ the concentration, and $x=r/r_{\rm vir}$. It is well known that the presence of an important stellar component can lead to a contraction of the dark matter \citep[e.g.][]{Blumenthal:1985qy,TrujilloGomez:2010yh}. Adiabatic contraction (AC) can be modelled as
\begin{equation}
\frac{r_f}{r_i}-1=\xi\left[\frac{v^2_{\rm dm}(r_i)}{(1-f_{\rm star})v_{\rm dm}^2(r_i)+v_{\rm star}^2(r_f)}\frac{r_i}{r_f}-1\right]
\end{equation}
where $r_i$ and $r_f$ are the radii of the initial and final mass bins. Due to analytical arguments, the parameter $\xi$ was originally assumed to be unity, corresponding to the case of angular momentum conservation at every radius \citep{Blumenthal:1985qy}. Later on it was realised that a value of $\xi=0.68$ provides a better match to simulations \citep{Gnedin:2004cx}. In this paper we investigate three cases of adiabatic contraction: $\xi=0$ (no AC),  $\xi=0.68$ (average AC), and  $\xi=1$ (strong AC). This allows us to conservatively bracket the minimum and maximum effect on the value of $v_{\rm max}$.

In Fig.~\ref{fig:velbias} we show the ratio $v_{\rm max}/v_{\rm max,dmo}$ where $v_{\rm max,dmo}$ is the initial dark-matter-only maximum circular velocity without assuming a stellar component. The three cases bracketing the full range of the AC effect are plotted in cyan, magenta, and purple colours while the dashed and dotted lines correspond to {\tt Planck} and {\tt WMAP3} cosmologies. We observe a mild cosmology dependence coming from the abundance matching procedure which is used to obtain the stellar-to-halo mass relation.

The increase of $v_{\rm max}$ due to the stellar disk is negligible below $v_{\rm max}\sim80$ km/s, has a prominent peak at $v_{\rm max}\sim130$ km/s and decreases again at $v_{\rm max}>200$ km/s. This is easily explained by the fact that Milky-Way sized galaxies have the largest stellar-to-halo mass ratios while both dwarfs and clusters are more DM dominated. We model the velocity bias with the equation
\begin{equation}\label{fitvelbias}
v_{\rm max} = \left[\frac{\alpha y^{\beta}}{1+y^{\beta+2}}+1\right]v_{\rm max,dmo},\hspace{0.5cm}y=\frac{v_{\rm max,dmo}}{v_{\alpha}}
\end{equation}
where $\beta=6$ and $v_{\alpha}=125$ km/s (see also Eq.~\ref{vmaxbias} of Sec~\ref{velbias}). For the remaining parameter $\alpha$, we assume three different values $\alpha=0.2,0.4,0.5$ where $\alpha=0.4$ corresponds to the best guess (assuming AC with $\xi=0.68$) while $\alpha=[0.2,0.5]$ bracket the minimum and maximum effect (see  solid lines in Fig.~\ref{fig:velbias}).

\section{Comparing our analysis to Read17}\label{app:read17}
In Sec.~\ref{VF}, the maximum circular velocities obtained with our simplified fitting approach are compared to the results from Read17 \citep{Read:2016aaa} and the good general agreement between the two is interpreted as a confirmation of our method. Here we take a closer look by performing an object-by-object comparison, and we discuss some of the differences between the two methods.

The results from Read17 are based on a re-analysis of the high-inclination galaxies from {\tt LITTLE THINGS} \citep{Oh:2015xoa} directly performed on the three-dimensional data cubes \citep{Iorio:2017aaa}. Read17 used the dark matter profile from \citet{Read:2015sta} to fit the full rotation curves obtained by \citet{Iorio:2017aaa}. The profile consist of an NFW function combined with a central core describing potential feedback effects from baryons, and it comes with three fitting parameters (one parameter for the core and two parameters for the NFW profile).

In contrast to the analysis from Read17, we only use galaxies with kinematic HI measurements that extend further out than three times the stellar half-light radius, thereby avoiding to deal with the poorly understood baryonic effects in the centres of galaxies. We furthermore restrict ourselves to the outermost data point of the velocity profile, even for galaxies where more data points are available. This allows us to keep the analysis pipeline as simple as possible and to obtain a general consistency between all analysed objects. However, at the scale of dwarf galaxies (i.e. the crucial scale for the present work), the large majority of our galaxies come from either the {\tt FIGGS} \citep{Begum:2008xc,Begum:2008gn} or the {\tt SHIELD} \citep{Cannon:2011fe} sample which only have one published data point. 
\begin{figure}
\center{
\adjustbox{trim={.03\width} {0.02\height} {0.02\width} {0.01\height},clip}{\includegraphics[width=.48\textwidth]{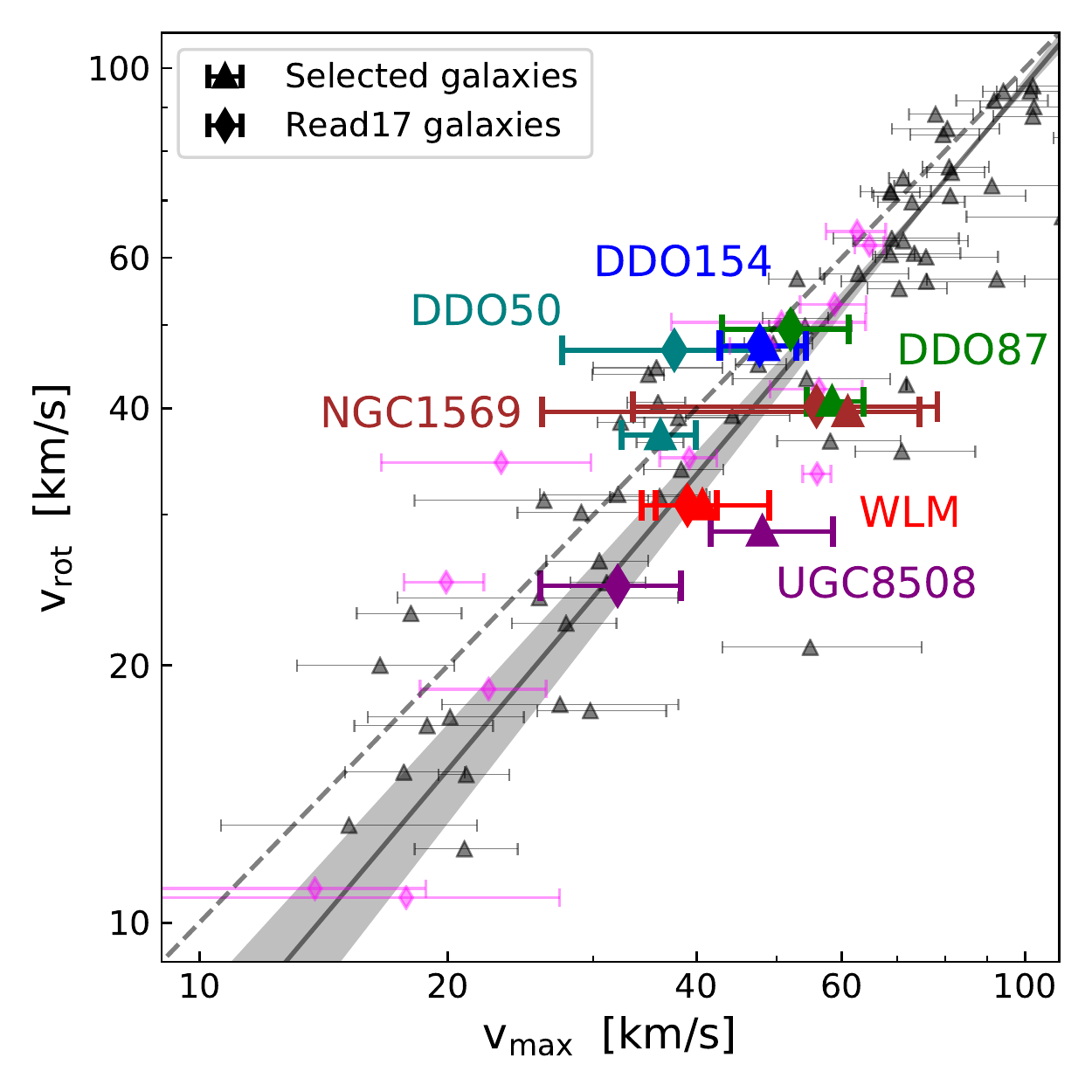}}
\caption{\label{fig:overlapp}Estimates of $v_{\rm max}$ from our analysis compared to Read17. The six overlapping galaxies analysed by both Read17 and us are shown as bold coloured diamonds and triangles, respectively. All remaining galaxies from the two samples are shown as grey and light magenta symbols. The agreement of $v_{\rm max}$ is very good except for \emph{UGC8508} where Read17 uses data from {\tt LITTLE THINGS} while we rely on data from {\tt FIGGS}. Two other galaxies (\emph{DDO50}, and \emph{DDO87}) show differences in $v_{\rm rot}$ due to different inclination estimates (see text).}}
\end{figure}

From the 19 galaxies analysed by Read17 only six also appear in our selected galaxy sample, while all others do not pass our selection criterion. In Fig.~\ref{fig:Planck} we show these six galaxies and compare their values of $v_{\rm rot}$ and $v_{\rm max}$ obtained by Read17 (coloured diamonds) and by our own analysis (coloured triangles). All galaxies are labelled with their name (same colour).

In terms of $v_{\rm max}$, five of the six galaxies show very good agreement, the estimated values from Read17 lying well within the error bars of our analysis. The sixth galaxy (\emph{UGC8508}), however, has a significantly larger value of $v_{\rm max}$ compared to the result from Read17. The reason for this discrepancy is that for \emph{UGC8508} we rely on data from {\tt FIGGS} while Read17 uses data from {\tt LITTLE THINGS}. We have checked that replacing the {\tt FIGGS} with the {\tt LITTLE THINGS} data in our analysis results in the galaxy \emph{UGC8508} to drop out of the selected sample as it fails to pass the selection criterion. 

In terms of $v_{\rm rot}$, two additional galaxies (\emph{DDO50} and \emph{DDO87}) show visible differences between our analysis and the one from Read17. This might surprise at the first glance since both measurements rely on the observational data from {\tt LITTLE THINGS}. However, a closer inspection shows that the origin of this discrepancy comes from different estimates of the galaxy inclinations. While we use the estimate from the original {\tt LITTLE THINGS} paper \citep{Oh:2015xoa}, Read17 relies on the more recent re-analysis of \citet{Iorio:2017aaa}.

In summary, the agreement between the $v_{\rm max}$ estimates of Read17 and our simplified method is very encouraging. It confirms that limited information about the HI velocity is sufficient to estimate the total halo mass as long as the measurement comes from the outermost parts of the gas disk far away from the stellar component.
\end{document}